\documentclass[fleqn,usenatbib]{mnras}

\usepackage{newtxtext,newtxmath}

\usepackage[T1]{fontenc}
\DeclareRobustCommand{\VAN}[3]{#2}
\let\VANthebibliography\thebibliography
\def\thebibliography{\DeclareRobustCommand{\VAN}[3]{##3}\VANthebibliography}

\usepackage{graphicx}	
\usepackage{amsmath}	
\usepackage{subcaption}
\usepackage[normalem]{ulem}

\title[Mini Mergers and AGN in IllustrisTNG]{Interacting galaxies in the IllustrisTNG simulations - IX: Mini mergers trigger AGN in cosmological simulations}

\author[S. Byrne-Mamahit et al.]{Shoshannah Byrne-Mamahit,$^{1}$\thanks{E-mail: sjbyrnem@uvic.ca}
Sara L. Ellison,$^{1}$
David R. Patton,$^{2}$
Scott Wilkinson,$^{1}$
\newauthor Leonardo Ferreira,$^{3,1}$
Connor Bottrell,$^{4}$
\\
$^{1}$ Department of Physics \& Astronomy, University of Victoria, 3800 Finnerty Road, Victoria, British Columbia, V8P 5C2, Canada\\
$^{2}$ Department of Physics \& Astronomy, Trent University, 1600 West Bank Drive, Peterborough, Ontario, K9L 0G2, Canada\\
$^{3}$ Instituto de Matemática Estatística e Física, Universidade Federal do Rio Grande, Rio Grande, RS, Brazil\\
$^{4}$ International Centre for Radio Astronomy Research, University of Western Australia, 35 Stirling Hwy, Crawley, WA 6009, Australia}

\date{Accepted XXX. Received YYY; in original form ZZZ}

\pubyear{2025}

\begin{document}
\label{firstpage}
\pagerange{\pageref{firstpage}--\pageref{lastpage}}
\maketitle

\begin{abstract}
Galaxy mergers are transformative events that can cause gaseous inflows capable of triggering active galactic nuclei (AGN). Previous studies of AGN in simulations have mainly focused on major interactions (i.e. between approximately equal mass galaxies), which produce the strongest inflows and, therefore, would be the most likely to trigger AGN activity. However, minor interactions are far more common and may still enhance accretion onto supermassive black holes. We present an analysis of post-merger galaxies from the IllustrisTNG simulation with stellar mass ratios of $\mu>$1:100. We select post-mergers from the TNG50-1 simulation, from redshifts $0\leq z< 1$, with stellar masses greater than $10^{10}M_{\odot}$. We find an AGN excess in post-mergers with a stellar mass ratio as low as 1:40. The AGN excess is mass ratio and luminosity dependent, with 1.2-1.6 times more AGN found in post-mergers of 1:40$\leq \mu < $1:4 than in matched non-merger controls, and as many as 6 times more AGN found in major $\mu \geq$1:4 post-mergers. The AGN excess is long lived, between 500 Myr to 1 Gyr post-coalescence, across all of the mass ratio regimes. We demonstrate that the most luminous AGN in the simulation overwhelmingly occur in either post-mergers or pairs (with $\mu \geq $1:40). Finally, we demonstrate that mini mergers are likely to be overlooked in observational studies due to the weakness of features usually associated with recent merger activity, such as tidal streams and shells, making it challenging to completely account for merger-induced AGN activity even in deep galaxy surveys.
\end{abstract}



\section{Introduction}
\label{sec:introduction}

Our current picture of galaxy formation is predicated on hierarchical assembly. In the $\Lambda$CDM paradigm, all massive galaxies must form their dark matter halos through mergers with smaller substructures, which cements galaxy mergers as both a ubiquitous and critical component in galaxy evolution \citep{WhiteReese1978,Blumenthal1984}. Moreover, the dynamic interactions that occur during a galaxy merger are also predicted to have significant effects on the galactic properties. Binary simulation studies of high mass ratio (nearly equal mass) mergers suggest that the gravitational torques generated during the interaction are an efficient mechanism for draining angular momentum in the galaxy's gas, generating inflows to the galactic centre \citep{Hernquist1989a,Barnes1991,CapeloDotti2017,Blumenthal2018}. Simulations of isolated galaxy mergers predict that major mergers might play a significant role in galaxy evolution \citep{Hernquist1989a,Barnes1991,DiMatteo2007}, by increasing accretion onto the supermassive black hole (SMBH) \citep{DiMatteo2005,Springel2005,Hopkins2008,Capelo2015,Choi2024}, and even leading to a rapid truncation of star formation activity \citep{Hopkins2008}. Modern cosmological simulations corroborate some of the above predictions from binary simulations, finding enhanced star formation rates \citep{Patton2020,Hani2020,Faria2025,Schechter2025} and SMBH accretion rates \citep{Steinborn2018,McAlpine2020,ByrneMamahit2023,ByrneMamahit2024,Schechter2025}. However, cosmological simulations also diverge from the isolated merger simulation predictions. For example, the majority of mergers do not rapidly quench \citep{Quai2021,Quai2023}, although \cite{Davies2020} find that galaxies with a richer merger history are more likely to be quenched over long(er) timescales. 

Observations indicate that interacting galaxies, as a population, demonstrate enhancements in star formation rates \citep{Barton2000,Woods2006,Woods2007,Ellison2008,Woods2010,Scudder2012,Ellison2013,Patton2013,Knapen2015,Cao2016,Thorp2019,Bickley2022} and higher fractions of recently quenched galaxies \citep{Ellison2022,Li2023psb,Ellison2024}. However, there exists some tension in the observational literature surrounding mergers and AGN, likely complicated by a number of factors. Generally, AGN are found in excess in post-merger \citep{Ellison2013, Li2023} or paired galaxies \citep{Alonso2007,Ellison2011,Silverman2011}, where the strength of the AGN excess seems to be dependent on the stage in the merger sequence, with higher excesses found in the post-coalescence phase compared with the pair phase \citep{Satyapal2014,Bickley2023,Bickley2024a,Comerford2024,VasquezBustos2025}. There is a debate on the role of luminosity, with many studies finding an overabundance of disturbed morphologies in the most luminous AGN \citep{Bennert2008,Urrutia2008,Treister2012,Bessiere2012,Glikman2015,Hong2015,Pierce2022, LaMarca2024}. Conversely, other studies find normal incidences of merger features in AGN compared with non-AGN controls \citep{Cisternas2011,Kocevski2012,Bohm2013,Mechtley2016,Villforth2017,Marian2019}, and \cite{Villforth2023} perform a systematic study of merger fractions in various AGN samples from literature and conclude that there is no correlation with AGN luminosity. Additionally, redshift may play a factor, with some studies at higher redshift (z$\sim$2-3) finding no excess, or a weaker excess compared with lower redshifts, of AGN \citep{Shah2020,Silva2021,Dougherty2024}. Finally, one has to consider the non-trivial effects of AGN obscuration, with the highest merger excesses often found in mid-IR selected AGN \citep{Satyapal2014,Weston2017,Secrest2020,LaMarca2024,Ellison2025}. Despite the challenges, observations do broadly agree that mergers can trigger AGN (as evidenced by the excess of AGN observed in mergers compared with non-merger controls), but that most AGN do not appear to be mergers \citep{Schawinski2011,Schawinski2012,Villforth2014,Hewlett2017}.

Observations of galaxy mergers have, until recently, been limited in their capacity to study the merger sequence with temporal precision. In the pre-merger regime, one can use pair separation as a proxy for time (e.g. \citealt{Ellison2008,Silverman2011}), which does broadly separate early stage pre-mergers from those near coalescence \citep{Patton2024}. In contrast, without a reliable mapping between post-merger features and the time since coalescence, studies of post-merger populations present a time-averaged result which may include immediate and long-term post-mergers. However, recently, \cite{Ferreira2024,Ferreira2025} have constructed a machine vision model that is capable of not only identifying mergers, but also predicting their time since coalescence. \cite{Ferreira2025} and \cite{Ellison2025} have leveraged predictions of this merger classifier on deep r-band imaging to confirm the long-lived excess of star formation rate enhancements and AGN observed in post-mergers between 500 Myr to 1 Gyr after coalescence. These works demonstrated for the first time in observations that the effects of the merger event persist in the post-merger population up to a Gyr after the merger event. Such results emphasize that the timescale over which gravitational interactions influence the participating galaxies is significantly longer than just the immediate coalescence period when mergers are most identifiable. 

To summarize the complicated landscape of observational studies, mergers seem to host an overabundance of AGN and the enhancements of nuclear activity are long-lived. However, most AGN do not appear to be associated with merger activity. The above review of both the observational and theoretical merger literature shows that a wide variety of parameters, from redshift and luminosity dependence, through timescales and multi-wavelength perspectives, have been previously investigated. However, all of these works either explicitly (e.g. mass ratio cuts in pairs samples) or implicitly (e.g. post-merger samples based on visual or machine classifications: \citealt{Lotz2010b,Bickley2023}) select samples of mergers with mass ratios down to $\sim$1:10. Furthermore, many studies explicitly focus on the major regime (mass ratios $\geq$ 1:4) which is predicted to be associated with the strongest inflows; indeed \cite{Johansson2009} find a continuous decrease in the peak SF and SMBH activity from 1:1 mass ratio mergers down to 1:6 mass ratio mergers. However, minor mergers (1:10$\leq \mu <$1:4) are more common than major mergers \citep{Conselice2022}. Therefore, if minor interactions were capable of triggering an enhancement of AGN activity, their strength in numbers could allow them to contribute significantly to the AGN population. In fact, mergers in the minor regime (1:10 $\leq \mu $ < 1:4) have been demonstrated in simulations to trigger both star formation \citep{Cox2008,Hani2020} and AGN activity \citep{Capelo2015,Yang2019}. Furthermore, in \cite{ByrneMamahit2023}, we demonstrated that enhancements to the SMBH accretion rate were present in simulated post-mergers from the IllustrisTNG100-1 cosmological simulation, down to the lowest mass ratio investigated (1:10). 

There is nothing fundamental about a mass ratio of 1:10, yet little previous work has investigated the role of mergers more minor than this threshold. In the work presented here, we aim to extend the investigation of post-mergers and AGN down to the rarely explored mass ratio regime of mini mergers (1:100 $\leq \mu <$ 1:10). \cite{Bottrell2024} performed the first (to our knowledge) analysis of a complete population of post-mergers down to a mass ratio of 1:100 in simulations, and find that mini mergers play a significant role in modulating the scatter seen in the star forming main sequence, due to the presence of weaker but significant enhancements to galaxy star formation rate. In fact, \cite{Bottrell2024} find that mini mergers, although weaker, contribute more in integrated stellar mass excess, because they are much more common than their stronger but rarer major merger counterparts. Such results emphasize that a complete accounting of the effects of interactions on AGN triggering may require galaxy mergers of lower mass ratio than 1:10.

In this paper, we will explore a complete sample of post-mergers over a mass ratio range of 1:100 $\leq \mu <$ 1:1 from the IllustrisTNG50-1 cosmological simulation, and quantify their enhancements to the SMBH accretion rates. In Section \ref{sec:methods}, we describe the methodology for identifying mergers in IllustrisTNG, and outline our sample selection criteria. We investigate the SMBH accretion rates of the merger sample in Section \ref{subsec:minMR}, investigate the timescale of merger driven AGN excesses in Section \ref{subsec:timescale}, quantify the contribution of interacting galaxies to the overall AGN population in Section \ref{subsec:AGNpop}, and look at the visual features of mini mergers in Section \ref{subsec:whatdotheylooklike}. Finally, we discuss the effects of matching methodology in Section \ref{subsec: matching with Gas} and the effects of simulation resolution in Section \ref{subsec:resolution}, and summarize the conclusions of the work presented here in Section \ref{sec:conclusion}.

\section{Methods}
\label{sec:methods}

\subsection{The IllustrisTNG Simulation}
\label{subsec:TNG}

We use the magneto-hydrodynamic cosmological simulation IllustrisTNG \citep{Nelson2017,Pillepich20171,Springel2017,Mariancci2018,Naiman2018,Nelson2019,Nelson2019b,Pillepich2019}. The TNG simulations solve magneto-hydrodynamics with a moving-mesh approach using the \textsc{AREPO} magnetohydrodynamics code \citep{Springel2010}, which allows for adaptive spatial resolution, and gravity is solved using a tree-particle-mesh approach. We refer the reader to \cite{Weinberger2017} and \cite{Pillepich2018} for a full description of the galaxy physics models, and summarize here details pertaining to the SMBH physics models. 

SMBHs are seeded into dark matter halos exceeding $5 \times 10^{10} M_{\odot} h^{-1}$, at a seed mass of $8 \times 10^{5} M_{\odot}h^{-1}$. From here, SMBHs can increase in mass through accretion or SMBH-SMBH mergers. For SMBH accretion, TNG uses a Bondi-Hoyle-Lyttleton model, eq. \ref{eq:Bondi}, for parameterized spherically symmetric accretion onto a point mass \citep{Bondi1944,Bondi1952}
\begin{equation}
    \dot M_{Bondi} = \frac{4 \pi G^2 M_{BH}^2 \rho}{c_s^3},
    \label{eq:Bondi}
\end{equation}
where $G$ is the gravitational constant, $M_{BH}$ is the black hole mass, and $\rho$ and $c_s$ are the gas density and sound speed sampled in a kernel-weighted sphere enclosing $\sim $128 cells, centred on the SMBH. Accretion is Eddington limited, with a maximum permitted accretion rate of 
\begin{equation}
    \dot M_{Edd} = \frac{4 \pi G M_{BH} m_p}{\epsilon_r \sigma_T c},
    \label{eq:Edd}
\end{equation}
where $m_p$ is the proton mass, $\epsilon_r$ is the radiative accretion efficiency, $\sigma_T$ is the Thompson cross-section, and $c$ is the vacuum speed of light. The radiative accretion efficiency is taken to be 0.2 in TNG. When two SMBHs are within a kernel-weighted-sphere distance of one another, they are merged (combined into a single SMBH particle). Additionally, SMBHs are repositioned to the local gravitational minimum to avoid them wandering away from the galactic centre. For this reason, SMBH mergers often occur at the early stages of coalescence \citep{Bahe2022,Buttigieg2025}. 

In the simulation, gas temperature is affected by metal line cooling combined with an ionizing UV background and radiation heating from nearby AGN. AGN feedback occurs via two channels, a thermal heating (radiative) model at high SMBH accretion rates and a kinetic model at low accretion rates, determined following eq. \ref{eq:chi}
\begin{equation}
    \dot M_{BH}^{high} \geq \chi \dot M_{Edd} \hspace{0.5cm},\hspace{0.5cm} \chi = min\bigg[\chi_0 \bigg(\frac{M_{BH}}{10^8 M_{\odot}}\bigg)^{\beta},0.1\bigg],
    \label{eq:chi}
\end{equation}
where $\chi_0$ and $\beta$ are simulation parameters tuned to 0.002 and 2 respectively, and $\chi$ is capped to 0.1, following observational constraints set by X-ray binaries \citep{Dunn2010}. At high accretion rates, a fraction of the mass accretion energy is converted into thermal energy which is redistributed into the gas surrounding the SMBH. At low accretion rates, the mass accretion energy contributes to a reservoir which is eventually released in the form of kinetic `kicks' to the gas surrounding the SMBH.

For the work presented here, we use the TNG50-1 simulation box \citep{Nelson2019,Pillepich2019}, the highest resolution TNG simulation with a 50 cMpc simulation box size. We use TNG50-1 for our study of $\mu >$ 1:100 mergers as it allows us to probe well resolved (>1000 particles) galaxies in the simulation down to a stellar mass of $10^{8} M_{\odot}$. The dark matter mass resolution for the simulation is $4.5 \times 10^{5} M_{\odot}$ and the target baryon mass resolution is $8.5 \times 10^{4} M_{\odot}$. The minimum gravitational softening length for gas cells is 74 pc and 288 pc for dark matter particles. For a discussion on the effects of simulation resolution, see Section \ref{subsec:resolution} where we compare our results with the lower resolution boxes: TNG50-2 (dark matter mass resolution of $3.6 \times 10^{6} M_{\odot}$), and TNG100-1 (dark matter mass resolution of $7.5 \times 10^{6} M_{\odot}$).

\subsection{Merger and AGN Identification}
\label{subsec:mergerID}

\begin{figure}
    \centering
	\includegraphics[width=\columnwidth]{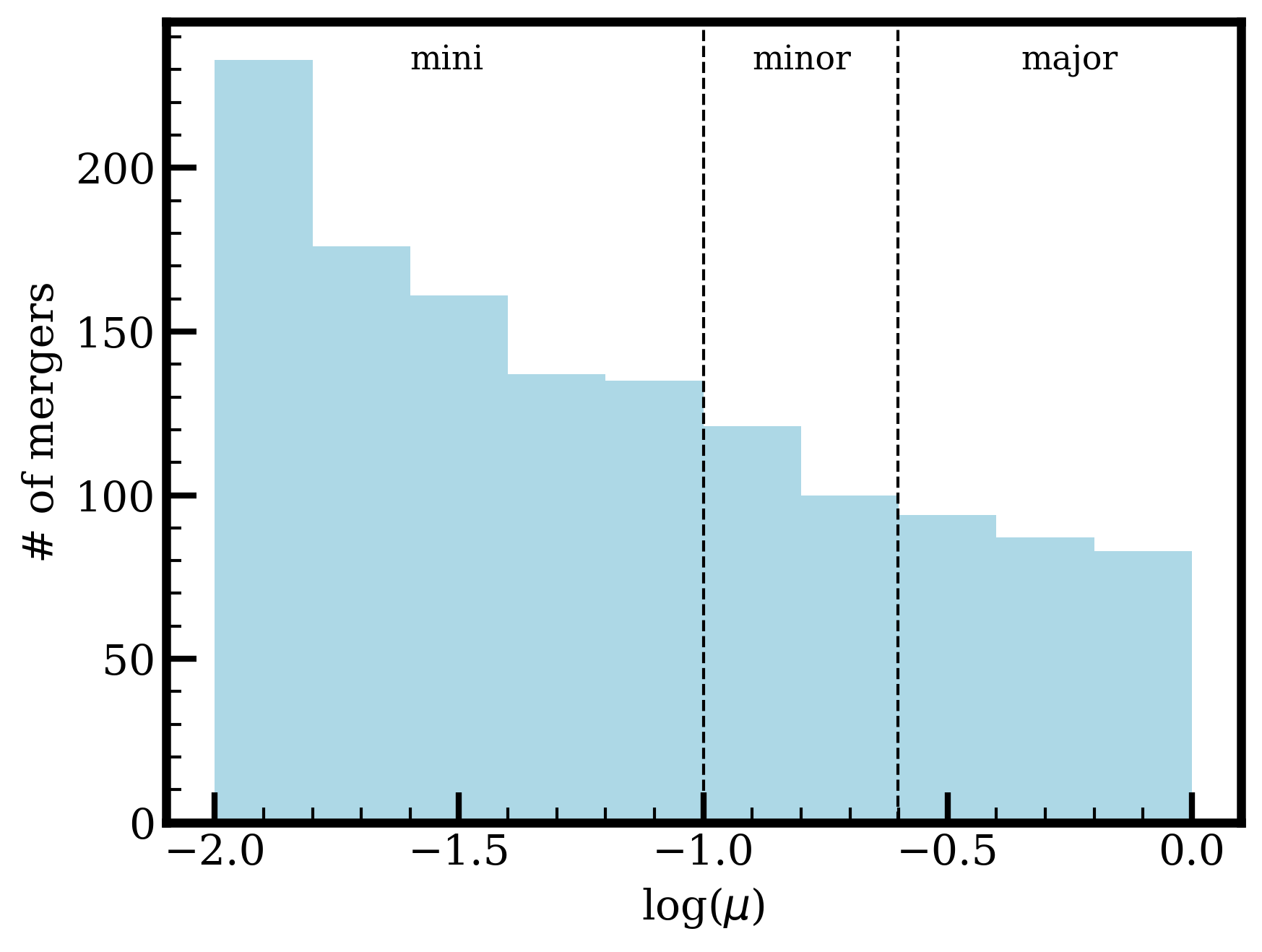}
    \caption{The distribution of the TNG50-1 post-merger mass ratios for mergers meeting the criteria of: a stellar mass of at least $10^{10} M_{\odot}$, z<1, and the merger is between subhalos originating from separate FOF groups. The vertical dashed lines delineate the three mass ratio regimes considered in this paper, namely major ($\mu \geq$ 1:4), minor (1:10 $\leq \mu <$ 1:4) and mini (1:100 $\leq \mu <$ 1:10).}
    \label{fig:MergerMassRatios}
\end{figure}

We identify galaxy mergers in the TNG50-1 simulation using merger trees created from the  \textsc{SubLink} algorithm \citep{Rodriguez-Gomez2015}. Briefly, \textsc{SubLink} merger trees link progenitor and descendant subhalos using the dark matter particles in the simulation. The algorithm compares the occupation of dark matter particles in subhalos between adjacent snapshots, creating a hierarchical `tree' tracing the assembly history of a subhalo. If multiple galaxies at snapshot N share a descendant galaxy at snapshot N+1, we identify this as a candidate merger. 

For each candidate merger, we assess the following properties: the \texttt{SubhaloFlag}, the length of the merger `branch', and the merger mass ratio. Subhalos that are likely to be of `non-cosmological origin' are identified using the following criteria and assigned a \texttt{SubhaloFlag=False}: they are satellites at the time of formation, they form within the virial radii of the parent halo, and the ratio of dark matter to total mass is less than 0.8 at the time of formation. For the work presented here, we limit our sample only to subhalos for which \texttt{SubhaloFlag=True}. When investigating candidate mergers, the galaxy `branch' length refers to the maximum amount of snapshots for which the \textsc{SubLink} algorithm can trace the galaxy's progenitors. We require a `branch' length of at least 5 snapshots in order that we may accurately measure the mass ratio according to the properties of the progenitor galaxies. 

Despite being a seemingly simple parameter, there is no single way (and no single \textit{right} way) to define the mass ratio of a merger. At the most basic level, the mass ratio is expected to evolve during the pair phase due to new stellar mass that is built up due to triggered star formation \citep{Barton2000,Ellison2008,Woods2010,Scudder2012,Patton2013}. In addition, there are further effects to consider, such as physical stripping of mass as the galaxy pair interacts or numerical stripping due to the allocation of particles into the incorrect subhalo during close encounters \citep{Rodriguez-Gomez2015}. These subtleties have led various authors to define mass ratios in a variety of ways such as measuring the mass ratio when the less massive galaxy reaches a maximum mass \citep{SotilloRamos2022,Bottrell2024} or using the maximum masses of the galaxies within the past 500 Myr \citep{Patton2020,Hani2020}.

We considered multiple mass ratio calculations, however we will only describe our preferred method for this study. We calculate the stellar mass ratio of the progenitor galaxies at the snapshot of first infall. Numerically, first infall corresponds to the first instance when the two galaxies are members of the same Friends-of-Friends (FoF) group. Consequently, we exclude any mergers between galaxies that have always belonged to the same FoF group throughout their merger tree history, which corresponds to the rejection of less than 4\% of our sample. We prefer the first infall methodology over some previously used definitions because the severity and timescale of numerical and physical stripping throughout the pre-merger sequence is very different for major mergers (1:1) compared with our smallest mass ratio mergers (1:100). Indeed, we find that other methods for measuring the mass ratio provide inconsistent results on either end of the mass ratio regime of the sample. By assigning the mass ratio at first infall, before any numerical or physical stripping may occur, we limit the bias introduced towards a preferred end of the mass ratio regime, while acknowledging that in some cases, the mass ratio assigned to the merger may significantly vary between the time of first infall and coalescence, due to subsequent star formation in either galaxy in the pair, or physical stripping due to the encounter. 

We limit the mass ratio of the mergers in this study to a minimum of 0.01, or 1:100, and the total subhalo stellar mass of the post-mergers to at least $10^{10} M_{\odot}$. The combination of these limits ensures that the minimum mass of a progenitor galaxy in a 1:100 merger is at least $10^{8} M_{\odot}$, corresponding to approximately 1000 particles for the given average baryon mass resolution. We also limit our sample to galaxies with a redshift of less than one. Applying the above criteria, we have a sample of 1327 post-mergers (selected in the snapshot immediately after coalescence). Figure \ref{fig:MergerMassRatios} shows the sample of 1327 post-mergers as a function of their mass ratio. We demarcate three mass ratio regimes which we will refer to in the latter sections: the mini mergers from 1:100 $\leq \mu <$ 1:10, the minor mergers from 1:10 $\leq \mu <$ 1:4, and the major mergers from 1:4 $\leq \mu <$ 1:1.

\begin{figure}
    \centering
	\includegraphics[width=\columnwidth]{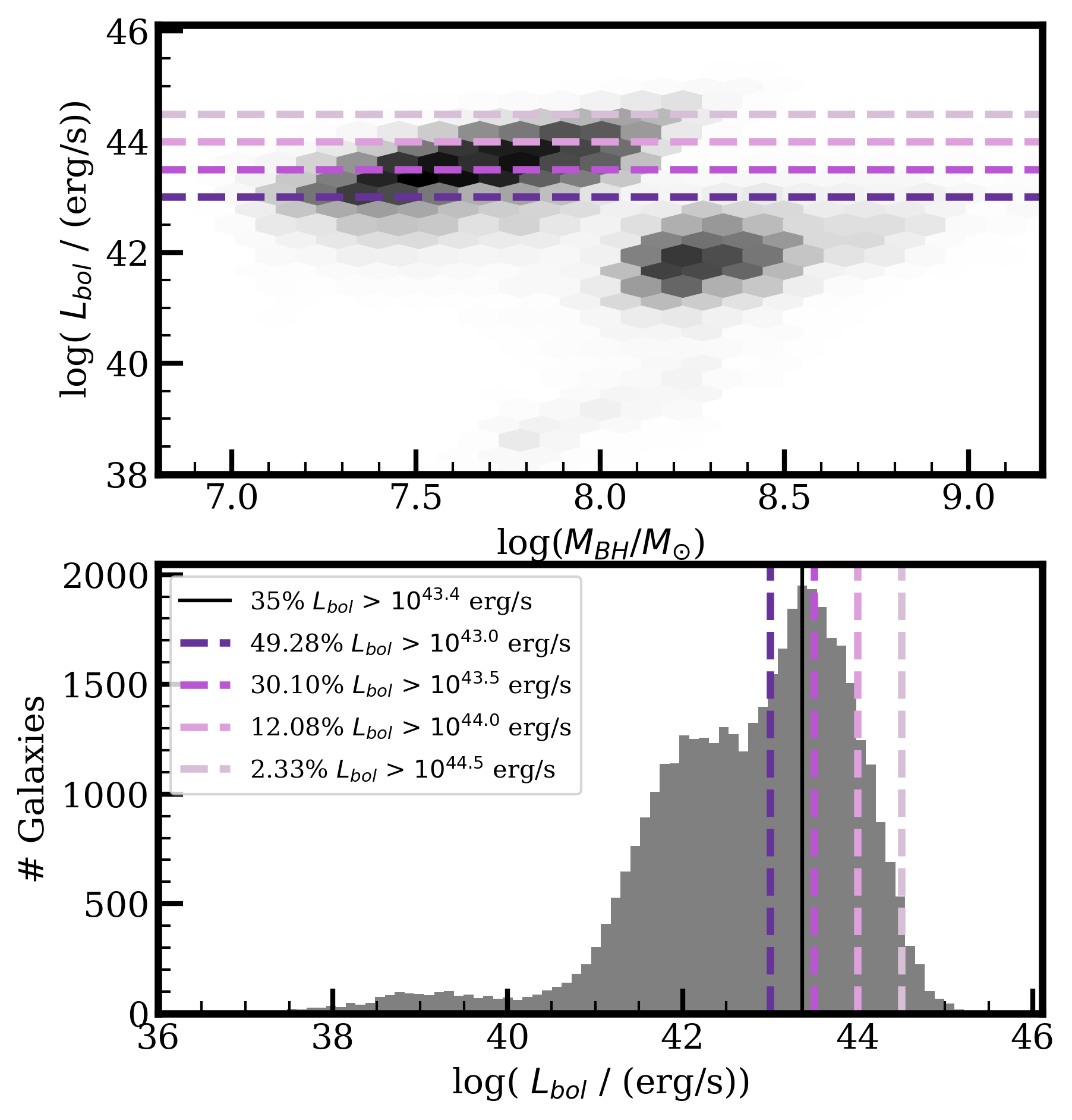}
    \caption{Top panel: the SMBH masses and luminosities for all TNG50-1 galaxies meeting our redshift and stellar mass criteria, i.e. stellar mass over $10^{10} M_{\odot}$ and $z<1$. Bottom panel: the luminosity function for galaxies in our sample. Different AGN luminosity regimes are highlighted in the dashed lines, and the percentage contribution to the overall population is shown in the legend. The most common (mode) accretion rate is shown in the solid line.}
    \label{fig:MainResult5.1}
\end{figure}

Finally, for the work presented here, we investigate AGN triggering by identifying the most highly accreting SMBHs in the simulation. We calculate the bolometric AGN luminosity as 10\% of the accretion mass energy, $L_{\text{bol}} = 0.1 \times \dot M_{BH}c^2$, for all galaxies in our mass and redshift limits, $M_{\star} \geq 10^{10} M_{\odot}$ and $z<1$. In this way, a SMBH will appear in the luminosity function at multiple points along its lifetime within the redshift limits of $z<1$. Figure \ref{fig:MainResult5.1} shows the distribution of AGN luminosities for all galaxies within our stellar mass and redshift limits. The top panel shows the distribution of SMBH masses and SMBH accretion rates (converted to bolometric luminosities) for galaxies within our mass and redshift limits. We see from the background density distribution that SMBHs are separated into two clouds, corresponding to high and low accretion modes. We define a range of AGN luminosity regimes, shown by the dashed horizontal lines. In the bottom panel, we show the luminosity function for the galaxies within our mass and redshift limits, where the solid vertical line shows the most common accretion rate for our sample. The legend displays the statistics for each luminosity regime, with just under 50\% of galaxies occupying our lowest luminosity regime of $L_{\text{bol}}>10^{43}$ erg/s, and only 2\% of galaxies occupying our highest luminosity regime of $L_{\text{bol}}>10^{44.5}$ erg/s.

\subsection{Matching Non-Merger Controls to Post-Merger Galaxies}
\label{subsec:Control Matching}

For each galaxy post-merger, we identify non-merger controls against which we may compare the SMBH properties. In order to identify non-merger controls, we begin by limiting the control pool to galaxies in TNG50-1 with a minimum total subhalo stellar mass of $10^{9} M_{\odot}$ and at redshifts less than one. We then further require that the galaxy has not had a merger of mass ratio $\mu \geq$ 1:100 within the past 2 Gyr, yielding a parent sample of 86,560 galaxies meeting the above criteria.

Each post-merger is control-matched only to galaxies in the same simulation snapshot, which serves as an exact match in redshift. We further restrict the control pool to galaxies with the same AGN feedback mode. We apply feedback mode matching following \cite{ByrneMamahit2023}, who demonstrated that, for our galaxy mass distribution, SMBH accretion rates follow a bimodal distribution separated by feedback mode. The bimodal distribution in SMBH accretion rates is seen in the top panel of Figure \ref{fig:MainResult5.1}. Within our mass and redshift limits, we find that SMBH accretion rates have a strong bi-modality, whereby galaxies that transition into a SMBH mass regime where effective kinetic mode feedback begins rapidly evacuate their central gas resulting in lower accretion rates compared with galaxies which are still predominantly using radiative mode feedback (see \citealt{Terrazas2020} for an analysis of the transition into effective kinetic mode feedback). We choose to match on instantaneous feedback mode in order to remove the added complexity of SMBH accretion rate enhancements which are actually due to post-mergers or controls transitioning from the high to low SMBH accretion regimes.

Finally, we match the following properties: subhalo stellar mass within 2 times the half-mass radius, the distance to a nearest neighbour with a mass of at least 1:10 the host ($\mathrm{r}_1$), and the number of neighbours with a mass of at least 1:10 the host ($\mathrm{N}_2$) within 2 Mpc of the host galaxy. We begin by selecting galaxies within the control pool that match the post-merger stellar mass within a tolerance of 0.1 dex, $\mathrm{r}_1$ to within 0.1 dex, and $\mathrm{N}_2$ to within 0.1 dex. If the post-merger does not have at least 5 control candidates, the tolerance in stellar mass, $\mathrm{r}_1$, and $\mathrm{N}_2$ are increased by 0.05 dex, 0.1 dex, and 0.1 dex respectively, up to a maximum of 3 times. If there are still fewer than 5 control candidates, the post-merger is rejected from the sample. We then select the best 5 controls, corresponding to the controls which minimize the following weighting factor given in eq. \ref{eq:control}
\begin{equation}
    w_i = \prod_i \Big(1-\frac{\mathrm{abs}(x_i^{\mathrm{PM}}-x_i^{\mathrm{control}})}{\delta x_i}\Big),
    \label{eq:control}
\end{equation}
where $x_{i}^{\mathrm{PM}}$ is the post-merger property, $x_i^{\mathrm{control}}$ is the control pool galaxy property, and $\delta_i$ is the acceptable tolerance for the given quantity. We therefore have 5 control galaxies for each post-merger that is successfully control matched.

The number of controls per merger, and the match tolerances were selected such that stellar mass and environmental variables of the post-mergers and matched controls pass a Kolmogorov-Smirnov (KS) test to within a 95\% confidence interval, while maximizing the number of successfully matched post-mergers.

\begin{figure}
    \centering
	\includegraphics[width=0.98\columnwidth]{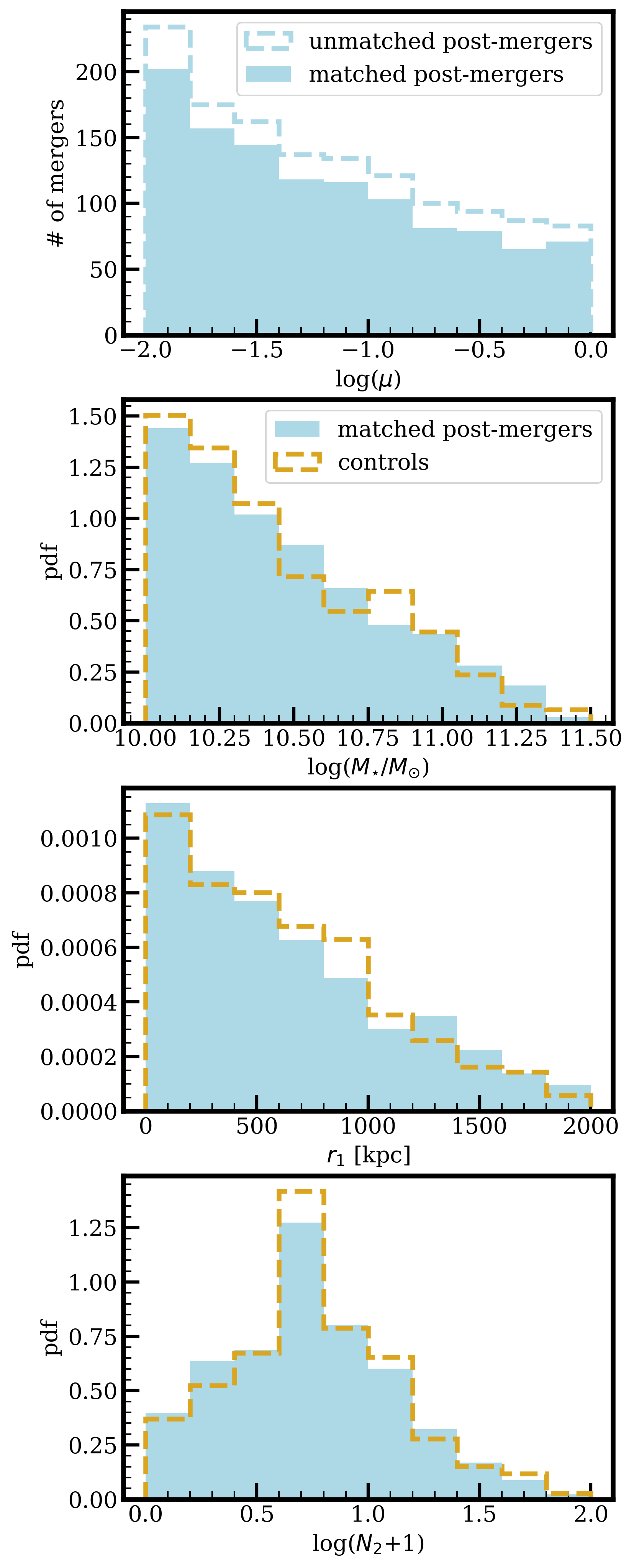}
    \caption{The distribution of the successfully control-matched post-mergers in mass ratio (first panel from the top), stellar mass (second panel), nearest neighbour distance $\mathrm{r}_1$ (third panel), and the number of neighbours $\mathrm{N}_2$ (fourth panel). In the first panel from the top, the dashed blue line shows the post-merger distribution before control matching. In the following panels, the dashed yellow line shows the distribution for the matched controls. In all panels, the matched post-mergers are shown in solid blue bars.}
    \label{fig:MatchedMergerSample}
\end{figure}

We reject 148 post-mergers which are not matched to a sufficient number of controls within our maximum error tolerances. Our final matched sample consists of 1179 post-mergers, each with 5 controls (for 5895 control galaxies). Figure \ref{fig:MatchedMergerSample} shows the results of our control matching procedure. In the top panel of the figure, we show the unmatched post-merger sample in the dashed blue line compared with the matched post-mergers in blue solid bars, demonstrating that our control matching procedure is not preferentially successful for a specific mass ratio regime. In the remaining panels we compare the matched post-mergers shown in the blue solid bars with their controls shown in the yellow dashed lines, demonstrating the quality of our control matching scheme.

\section{Results}
\label{sec:Results}

\subsection{What is the minimum mass ratio for SMBH accretion rate enhancements?}
\label{subsec:minMR}

\begin{figure}
    \centering
	\includegraphics[width=\columnwidth]{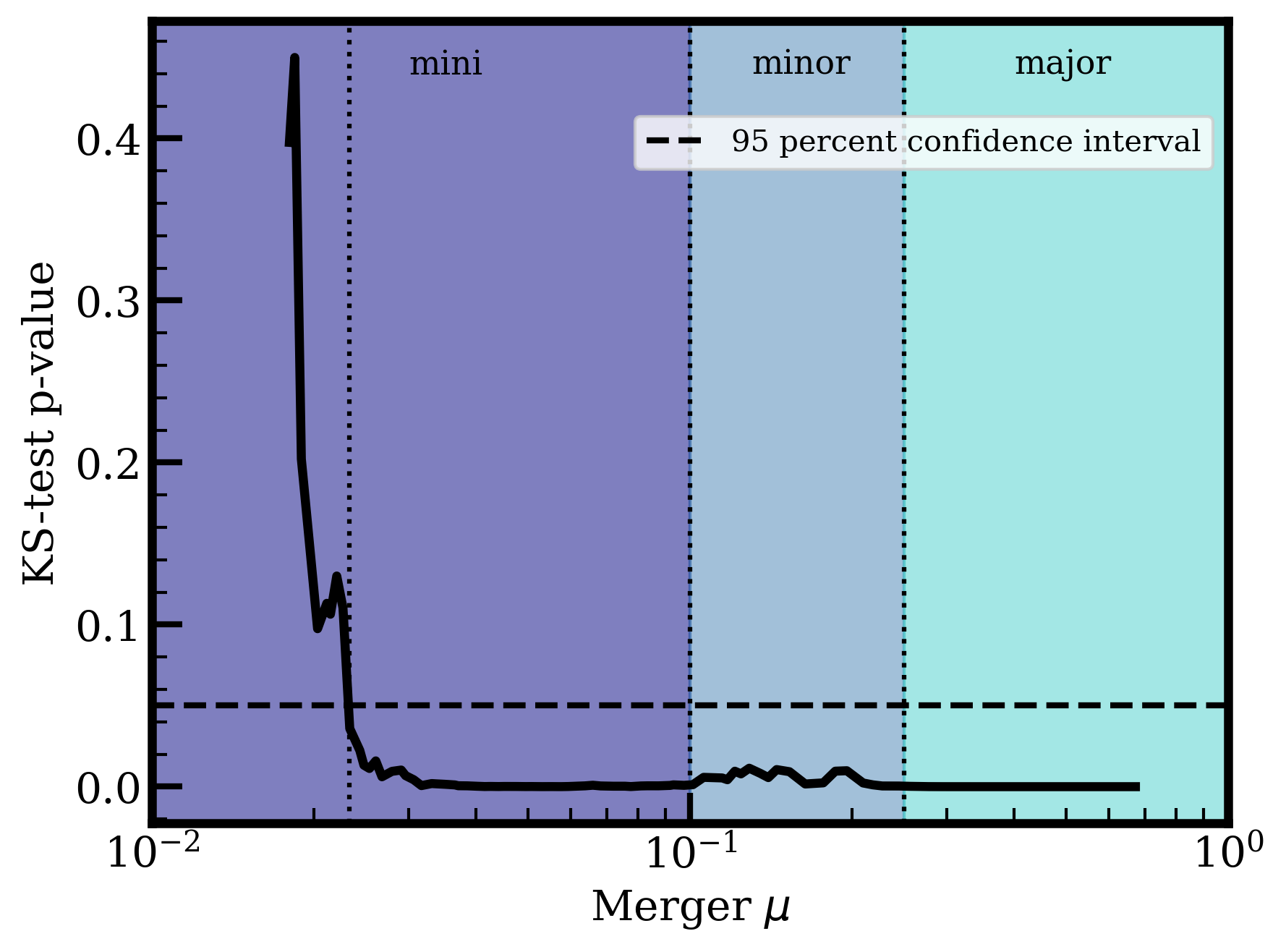}
    \caption{The probability that the distribution of the post-merger SMBH accretion rates (SMBHARs) are sampled from the same population as the matched control SMBH accretion rates, according to a KS-test. The KS-test is performed on post-mergers (and associated controls) with a mass ratio +/- 0.1 dex from the corresponding mass ratio of the x-axis. Post-mergers and controls have statistically indistinguishable SMBHARs at mass ratios below 1:40, and are statistically distinct for mass ratios above 1:40.}
    \label{fig:MainResult2}
\end{figure}

\begin{figure*}
    \centering
	\includegraphics[width=\textwidth]{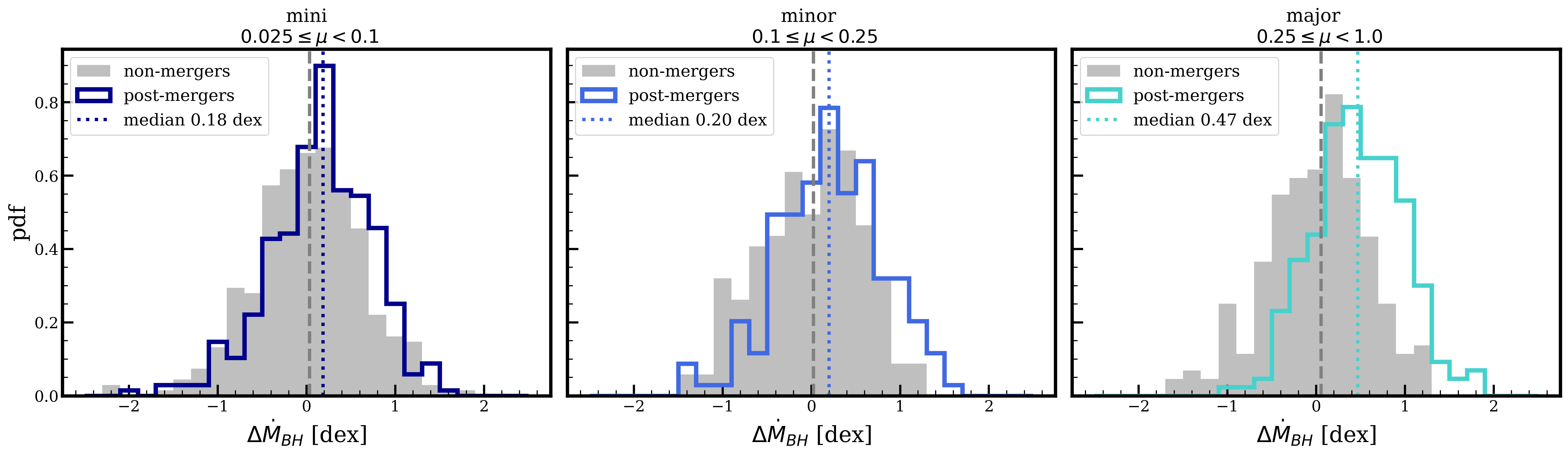}
    \caption{The offset of the post-merger SMBHAR compared with the median SMBHAR of the matched controls. Shown in grey is the expected width of $\Delta \dot M_{BH}$ due to SMBHAR stochasticity, demonstrated through a control matched non-merger population. Shown in blue are the post-mergers in three bins of mass ratio: mini mergers from 1:40 - 1:10 (left) , minor mergers from 1:10 - 1:4 (centre) , and major mergers 1:4 - 1:1 (right). All three populations show enhanced SMBHAR compared with controls, with the strongest enhancement found in major mergers.}
    \label{fig:MainResult1}
\end{figure*}

\begin{figure}
    \centering
	\includegraphics[width=\columnwidth]{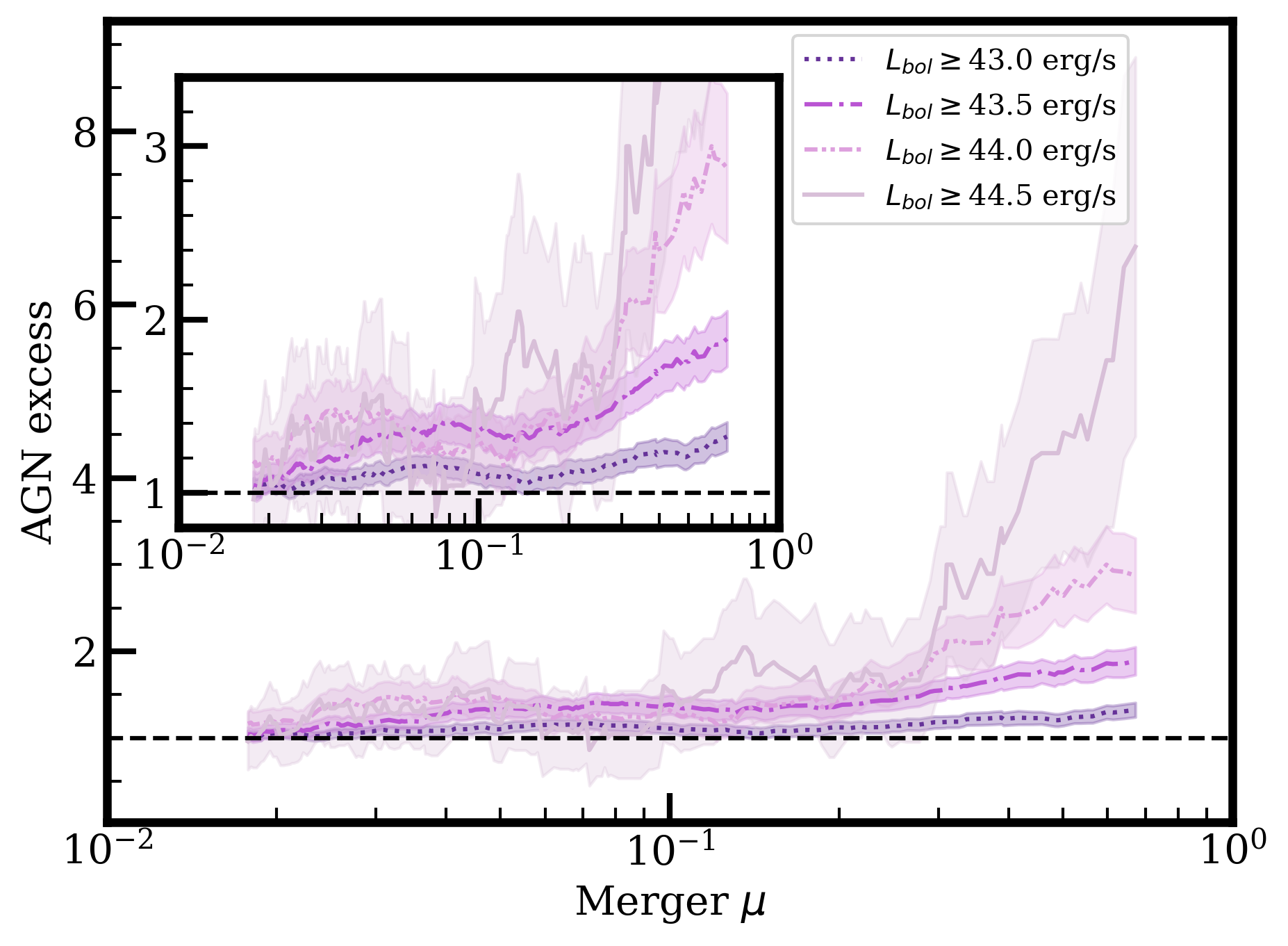}
    \caption{The excess of AGN (defined as exceeding a given $L_{\text{bol}}$ threshold) in post-mergers relative to non-merger controls vs merger mass ratio. The inset plot shows the same data zoomed over a smaller range on the y-axis. There is a statistically significant excess for the nearly all luminosity thresholds beginning at a mass ratio of 1:40 (excluding $L_{\text{bol}}>10^{44.5}$erg/s). The most luminous threshold only shows a statistically significant enhancement beginning at a mass ratio of 1:4.}
    \label{fig:MainResult3}
\end{figure}

Once the post-merger and matched non-merger control galaxies have been identified, we begin with the analysis of their SMBH accretion rates (SMBHAR). Figure \ref{fig:MainResult2} shows the result of a KS-test performed on the SMBHAR distribution of the post-mergers compared with their matched controls, as a function of merger mass ratio. In the figure, we demarcate the following mass ratio regimes: mini mergers from 1:100 $\leq \mu <$ 1:10 shown in dark blue, minor mergers from 1:10 $\leq\ \mu <$ 1:4 shown in medium blue, and major mergers from 1:4 $\leq \mu <$ 1:1 shown in light blue. The statistical test demonstrates that the SMBHAR distributions are statistically indistinguishable for mass ratios less than approximately 0.025 (1:40). For mass ratios greater than 1:40, they pass a 95\% confidence threshold as distinguishable populations. Our results therefore demonstrate a lower mass ratio limit of 1:40, below which SMBH accretion rates in post-mergers are indistinguishable from non-merging galaxies. We therefore adopt a lower boundary of 1:40 for the mini mergers used in our analysis, unless otherwise specified.

In addition to comparing post-merger and control SMBHAR distributions directly, we also investigate their SMBHAR enhancement (or deficit) using eq. \ref{eq:Delta}:
\begin{equation}
    \Delta \dot M_{\text{BH}} = log_{10}\Big(\dot M_{\text{BH}}^{\text{PM}} - \text{median}\big(\dot M_{\text{BH}}^{\text{controls}}\big)\Big)
    \label{eq:Delta}.
\end{equation} 
Therefore, if a post-merger has a $\Delta \dot M_{BH} > 0$ dex, the accretion rate of the post-merger is enhanced relative to its matched non-merger controls. For the following experiments, we separate the post-merger sample into mini (1:40-1:10), minor (1:10-1:4), and major (1:4-1:1) categories, i.e. our original definition of mini mergers for the following analysis has a truncated lower mass ratio bound from the original sample, since we have found (Fig \ref{fig:MainResult2}) that mass ratios below 1:40 do not alter the SMBHARs. Removing mergers of mass ratio less than 1:40 reduces the merger samples size to 920 (before control matching). Furthermore, in order to assess the effects of minor and mini mergers without `contamination' from recent more major events, we apply an additional criterion to minor and mini post-mergers. Minor mergers must have had at least 1 Gyr pass since the most recent merger of mass ratio $\mu\geq$1:4. Similarly, mini mergers must have had at least 1 Gyr pass since the most recent merger of mass ratio $\mu\geq$1:10. This `clean' post-merger sample consists of 858 post-mergers, of which 727 are successfully control matched.

We compute $\Delta \dot M_{BH}$ for each post-merger, and plot their distributions in Figure \ref{fig:MainResult1}. Shown in solid lines are the three mass ratio categories: mini mergers (left), minor mergers (centre), and major mergers (right). The dotted vertical lines are the median $\Delta \dot M_{BH}$ for each of the post-merger samples. To provide a reference for how we expect $\Delta \dot M_{BH}$ to behave in a non-merger sample, we take the best-matched control for each post-merger and in turn control match each of these galaxies to 5 new non-merger controls. The result of the matched non-mergers is shown in the background grey histograms. The grey dashed line is the median of the non-merger sample. 

Figure \ref{fig:MainResult1} shows that both samples (post-mergers and comparative non-mergers) have a characteristically wide distribution in $\Delta \dot M_{BH}$ due to the large intrinsic scatter in SMBHAR. In this way, even non-merger galaxies can have SMBH accretion rates up to 100 times higher or lower than the median SMBHAR of their matched non-merger controls. Since there does not exist a tight sequence for SMBHAR (as is seen, for example, between SFR and $M_{\star}$), we do not place a strong emphasis on each individual galaxy's $\Delta \dot M_{BH}$, but instead look at how the median and distribution of $\Delta \dot M_{BH}$ varies between post-mergers and controls. For all of the non-merger samples, the distribution is symmetric about 0 (as expected for a well-matched, non-interacting population). For post-mergers in all three mass ratio regimes, we find that the distributions of $\Delta \dot M_{BH}$ are similarly wide compared with the non-mergers, however the distributions are asymmetrically distributed around 0. For both mini and minor mergers, the medians of the distribution are very similar, 0.18 dex for mini post-mergers and 0.20 dex for minor post-mergers. Our results suggest that the degree of SMBHAR enhancement does not significantly vary from mass ratios 1:40 to 1:4. Finally, for major mergers, the distribution is very skewed to positive enhancements, with a median of 0.47 dex in post-mergers. Therefore, we find that the post-merger population begins to show SMBHAR enhancement at mass ratios greater than 0.025 (1:40), and that the degree of enhancement is similar between the mini (1:40-1:10) and minor (1:10-1:4) populations, but is strongest in the major ($\mu$>1:4) mergers. 

While we have demonstrated that mini post-mergers have enhanced SMBHARs, enhanced accretion does not necessarily mean mini mergers contribute to an excess of the most \textit{luminous} AGN in the simulation. To address this, we now investigate whether post-mergers host an excess of luminous AGN compared with their matched controls. We compute the fraction of galaxies with a bolometric AGN luminosity greater than a given lower luminosity limit in both the post-merger sample and the matched control sample. We then compute an AGN excess as the fraction of AGN (identified as exceeding a given $L_{\text{bol}}$ threshold) in post-mergers divided by the fraction of AGN in their associated controls. Figure \ref{fig:MainResult3} shows the AGN excess as a function of post-merger mass ratio. Different AGN luminosity thresholds are shown in the different colours and line-styles. The inset figure zooms in over a smaller dynamic range on the y-axis to emphasize the trends in the lower three luminosity ranges. Figure \ref{fig:MainResult3} demonstrates that in the lower three AGN luminosity ranges, an AGN excess is present beginning at a mass ratio of 1:40, in agreement with the results of Figure \ref{fig:MainResult2}. These AGN excesses are between 1.2-1.4 up to a mass ratio of 1:10. For $L_{\text{bol}} \geq 10^{43.5}$ erg/s and $L_{\text{bol}} \geq 10^{44}$ erg/s, the excess increases up to 1.8 and 2.8 for the most major mergers. For the most luminous AGN, $L_{\text{bol}} \geq 10^{44.5}$ erg/s, a statistically significant excess is only present above mass ratios of approximately 1:4. For all luminosities, the AGN excess is most strongly correlated with mass ratio for mass ratios in excess of 1:4. The correlation with mass ratio in this regime is in agreement with the idealized simulation studies of \cite{Johansson2009}, who found that the peak SMBH accretion rates correlated with mass ratio in their simulated mergers of mass ratio 1:1 to 1:6. Additionally, the maximum excess of approximately 7 is found in the most luminous AGN in the most major mergers. Our results therefore highlight that intermediate luminosity AGN are present in excess in the post-merger population down to mass ratios of 1:40. However, the most luminous AGN are only found in (significant) excess in the major merger regime.

\subsection{How long do AGN excesses last in the post-merger population?}
\label{subsec:timescale}

\begin{figure*}
    \centering
	\includegraphics[width=\textwidth]{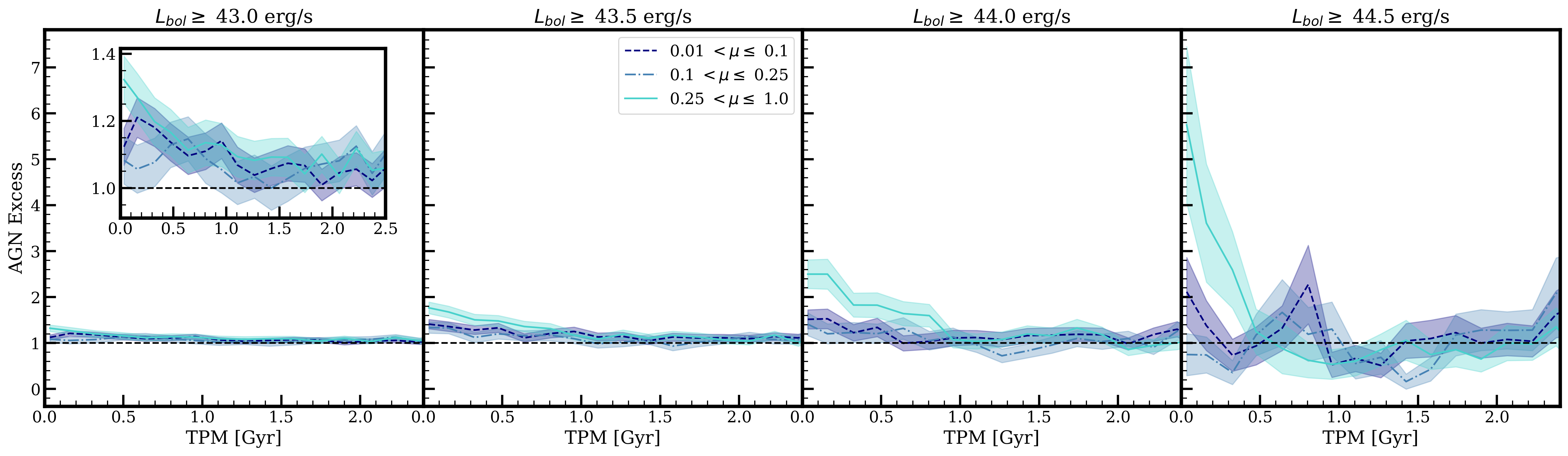}
    \caption{The AGN excess in post-mergers over the matched controls vs the time post-merger (TPM). The mini (1:40-1:10), minor (1:10-1:4), and major (1:4-1:1) mergers are shown in the dark dashed, medium dash-dot, and light solid blue lines respectively. The panels are organized from left to right in order of increasing AGN luminosity threshold. The AGN excess timescale decreases with increasing AGN luminosity, and is longer in major post-mergers compared with mini and minor post-mergers.}
    \label{fig:MainResult4}
\end{figure*}

Having determined that enhancements in SMBHAR appear to begin in mergers of mass ratio 1:40, we can now investigate the timescales of such enhancements and if there is a dependence on mass ratio and AGN luminosity. Figure \ref{fig:MainResult4} shows the AGN excess as a function of time post-merger (TPM). Each panel shows the luminosity threshold used to calculate the AGN excess. The mini (1:40-1:10), minor (1:10-1:4), and major (1:4-1:1) mergers are shown in the dark dashed, medium dash-dot, and light solid blue lines respectively. The y-scale is deliberately fixed for all four panels in order to demonstrate the luminosity dependence, but the inset in the first panel shows the AGN excess for the lowest luminosity limit with a smaller dynamic range on the y-axis.

Beginning with the lowest luminosity threshold (left-most panel of Fig \ref{fig:MainResult4}), we find that all three mass ratio regimes have a modest AGN excess immediately post-coalescence, by a factor of 1.1-1.3. In all three mass ratio regimes, the excess is long lived, returning to normal rates of low(er) luminosity AGN by 1 to 2 Gyr in the mini and minor post-mergers. However the major mergers maintain a very low level excess up to 2 to 2.5 Gyr post-merger. As we increase the luminosity threshold for the AGN selection, the divergence in behaviour of mini and minor post-mergers compared with major post-mergers becomes more pronounced. For the next two AGN luminosity regimes, $L_{\text{bol}}\geq 10^{43.5}$ erg/s and $L_{\text{bol}}\geq 10^{44}$ erg/s, the maximum excess at coalescence increases (up to 1.4-1.6 in mini and minor post-mergers and 2-3 in major post-mergers) with increasing AGN luminosity and the excess is higher in major mergers, as expected from Figure \ref{fig:MainResult3}. However, for AGN with $L_{\text{bol}}\geq 10^{44}$ erg/s, the timescale over which AGN excesses are present is shortest for mini post-mergers (500 Myr), and slightly longer for minor mergers (800 Myr) and longest in major mergers (1 Gyr). Finally in the most luminous ($L_{\text{bol}}\geq 10^{44.5}$ erg/s) AGN panel, while there are some bins of time post-merger with a significant excess of AGN in the mini post-mergers, there is no significant sustained AGN excess in the mini or minor post-mergers, consistent with the results of Figure \ref{fig:MainResult3}. Additionally, the timescale over which AGN excesses are present in major mergers is shortest for the highest luminosity threshold, around 500 Myr. Overall, regarding the timescale of AGN enhancements in post-mergers, we find a long-lived (500+ Myr) AGN excess in all mass ratio regimes and luminosities (excepting mini and minor post-mergers in the most luminous threshold). We also find that AGN excess timescales are longer in major mergers compared with mini and minor mergers.

We emphasize that the results of Figure \ref{fig:MainResult4} are not to be interpreted as the timescale of any single AGN event (which are likely to be short and stochastic, eg. \citealt{Blecha2018}), but rather the population-averaged occurrence of short AGN events. Therefore, the timescales determined here provide insight into how long after coalescence stochastic AGN events, enhanced by merger activity, can occur. The decrease in AGN excess timescale with increasing AGN luminosity could be interpreted as the connection between proximity to coalescence and the necessary conditions for fueling stochastic AGN events. Our results are consistent with a scenario where the conditions necessary for the most luminous AGN are most likely to occur near coalescence, and less likely to occur beyond 500 Myr. Whereas, for the intermediate luminosity AGN, later stage post-mergers may still have some residual central gas enhancement that is sufficient to drive excesses of moderate AGN. Alternatively, it may be a modest long lived excess in the most luminous AGN regime is obscured by low number statistics. That is, if high luminosity AGN are extremely rare, we may not have enough occurances to significantly capture a modest excess beyond 500 Myr.

The long lived (500 Myr - 1 Gyr) timescales of such enhancements are consistent with other simulation studies, which consider populations of post-mergers of mass ratios 1:10 and above \citep{McAlpine2020,Schechter2025}. In addition, the short(er) timescale of the most luminous AGN excesses in the population of major mergers agrees with \cite{Johansson2009} who find that the SMBH accretion rate steeply drops off after coalescence in their 1:1 major mergers. Looking to observations, we find that the long-lived AGN excess in the post-merger population is consistent with \cite{Ellison2025}, who used time-labelled post-mergers in the Ultraviolet Near- Infrared Optical Northern Survey (UNIONS) Survey to show that signatures for AGN can be seen for at least one Gyr after coalescence. We additionally find that the magnitudes of the enhancements in the most luminous bin are consistent with the AGN excesses found in \cite{Ellison2025}, however we caution that the luminosity regimes shown in this paper are not able to be directly compared with those in observations, and in fact \cite{Ellison2025} investigate moderate luminosity AGN while our sample consists of the top 2.3\% AGN from the TNG50-1 simulation. Interestingly, \cite{Ellison2025} find that the timescale of the population-averaged AGN excess is longer lived for more luminous infrared AGN, which is the opposite of the trend observed above. However, direct comparisons are challenging here due to potential selection biases when comparing dust obscured AGN to the total population of luminous AGN \citep{Blecha2018}.

\subsection{How much do mini, minor, and major mergers contribute to the AGN population?}
\label{subsec:AGNpop}

\begin{figure}
    \centering
	\includegraphics[width=\columnwidth]{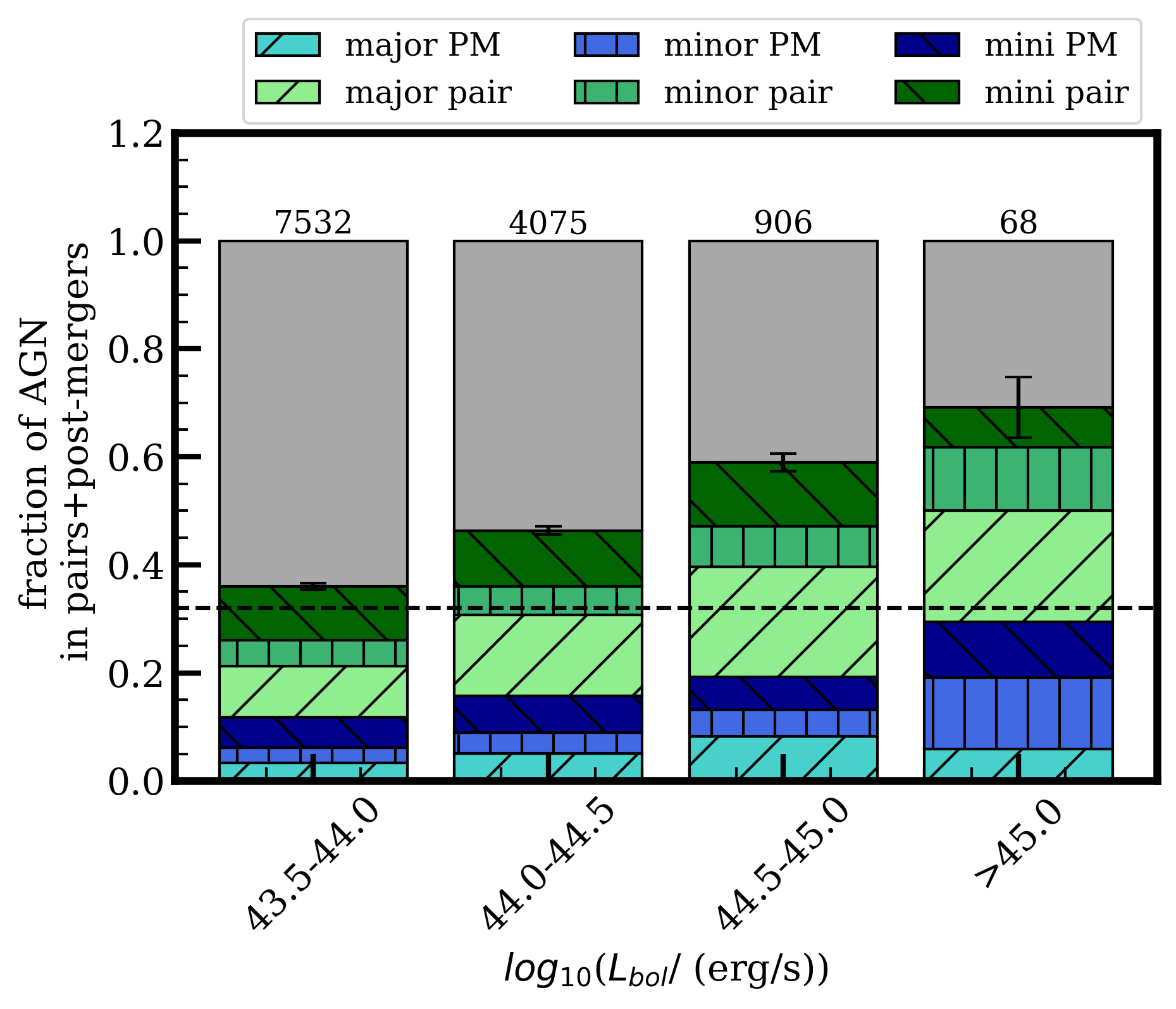}
    \caption{The fraction of highly accreting SMBHs in the categories of mini (1:40-1:10), minor(1:10-1:4) and major (1:4-1:1) pairs and post-mergers. Post-mergers are shown in shades of blue and pairs in shades of green, from darkest to lightest in order of increasing mass ratio. The numerical value above each bar indicates the number of galaxies in the AGN luminosity regime. The vertical error bar is the error on the interacting fraction, given by binomial statistics. The horizontal line is the interacting fraction for all galaxies of $M_{\star}>10^{10}M_{\odot}$ and z$<1$. The fraction of galaxies which are post-mergers or in pairs increases with increasing AGN luminosity, up to a majority fraction of $\sim70$\% for $L_{\text{bol}}>10^{45}\text{erg/s}.$}
    \label{fig:MainResult5.2}
\end{figure}

\begin{figure}
    \centering
	\includegraphics[width=\columnwidth]{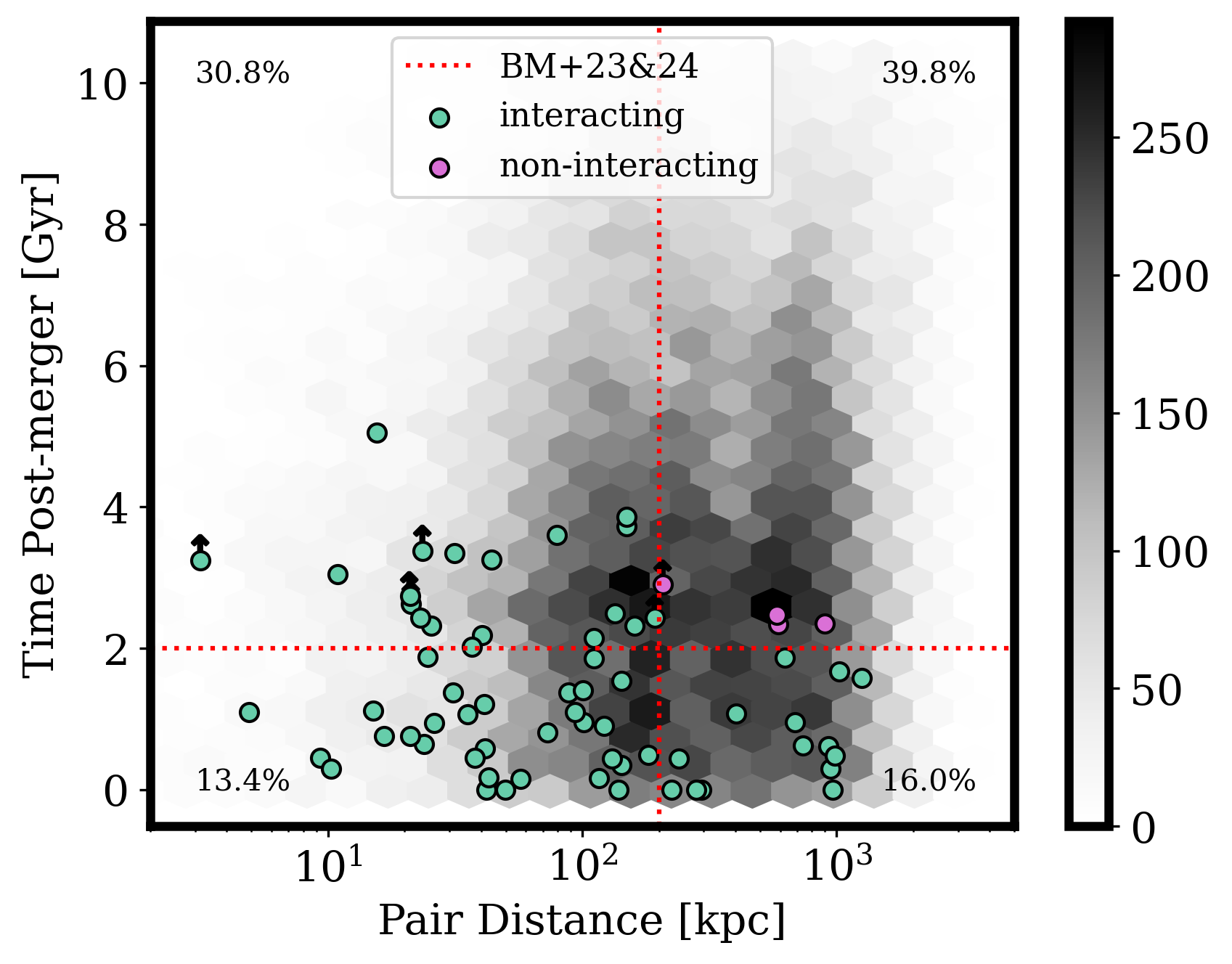}
    \caption{The distance to the nearest neighbour with a mass of at least 1/40th the host mass, and the time since the most recent merger event of $\mu\geq1:40$. The background density plot shows all galaxies above a stellar mass $10^{10} M_{\odot}$ and redshift < 1, with the occupation fraction for all galaxies shown in the corner of each quadrant. Coloured points are the galaxies from the most luminous ($L_{\text{bol}} \geq 10^{45}$ erg/s) AGN bin in Figure \ref{fig:MainResult5.2}. The red dotted lines correspond to the maximum distance (200 kpc) and time post-merger (2 Gyr) for which we see a significant AGN excess in \protect\cite{ByrneMamahit2023} and \protect\cite{ByrneMamahit2024}. Luminous AGN falling within these limits are shown in turquoise circles. Circles with arrows represent galaxies for which TPM is a lower limit (i.e. no merger as far as the merger tree can be followed or no merger up to z=2). The pink circles are the remaining galaxies which do not fall within the extended pair and post-merger limits. We find that the upper right quadrant (corresponding to the strictly non-interacting galaxies) is under-populated by luminous AGN compared with the overall population of galaxies (5.9\% vs. 39.8\%).}
    \label{fig:MainResult5.3}
\end{figure}

In the above sections, we consider the frequency of AGN triggering in the merger population. We now invert the question, and ask instead what the contribution is of mini, minor, and major mergers to the total luminous AGN population. In Figure \ref{fig:MainResult5.2}, we select all of the TNG50-1 galaxies meeting our redshift and stellar mass criteria, stellar mass over $10^{10} M_{\odot}$ and $z<1$, and separate them according to their AGN luminosity. Recall that Figure \ref{fig:MainResult5.1} provides statistics on the AGN regimes. For example, 30\% of our galaxy sample has $L_{\text{bol}} \geq 10^{43.5}$ erg/s, 12\% has $L_{\text{bol}} \geq 10^{44}$ erg/s, and only 2\% has $L_{\text{bol}} \geq 10^{44.5}$ erg/s.

For each luminosity bin in Figure \ref{fig:MainResult5.2}, we attribute the AGN galaxies into categories of post-mergers or pairs. Post-mergers are galaxies that have undergone a merger in the last 500 Myr, and pairs are galaxies that have a companion within 100 kpc (limits which we will re-visit later in the analysis). We separate both pairs and post-mergers into mini, minor, and major categories. Pair mass ratios are calculated following the procedure of \cite{Patton2020}, wherein for galaxy pairs that are significantly overlapping the mass ratio is evaluated using $M_{\star}^{\text{max}}$ (the maximum mass of the galaxy within the past 500 Myr). We then compute the distance to the nearest neighbour with a mass of at least 1/40th the host mass for the mini pairs, 1/10th the host mass for the minor pairs, and 1/4th the host mass for the major pairs. Galaxies are assigned into the categories with priority to post-mergers over pairs (in the case a galaxy is both a post-merger and a member of a pair), and with priority in order of major to minor to mini. For example, in the case where a galaxy is a mini post-merger within 100 Myr but also a major post-merger within 300 Myr, the galaxy is labelled a major post-merger. Finally, minor post-mergers are given priority over major pairs. While the priority order affects the specific categories, the overall interacting fraction (all pairs + all post-mergers) is unchanged. The interacting fraction of the total galaxy sample (all galaxies without any luminosity cuts) is shown in the black dashed line.

The lowest AGN luminosity bin in Figure \ref{fig:MainResult5.2} corresponds to a regime of accretion rates that are quite common in TNG50-1, and is similar to the expected contributions of each category in the overall galaxy population (black dashed line). We find that in total, all interacting categories account for 35 percent of the AGN population. The fraction of galaxies in interactions increases with increasing AGN luminosity, reaching a peak of 69 percent for our most luminous ($L_{\text{bol}} \geq 10^{45}$ erg/s) AGN bin. The trend of increasing interaction fraction with increasing AGN luminosity in TNG50-1 is consistent with what was previously demonstrated in TNG100-1 in \cite{ByrneMamahit2024}, further demonstrating that the most highly accreting SMBHs are predominantly found in galaxies that are in close pairs or post-mergers. However, in the work presented here, we have furthered the work of \cite{ByrneMamahit2024} by including a more complete census of interactions, extending our statistics down to the mini mass ratio regime. In addition, we find that in all AGN luminosity bins, the contribution of mini and minor post-mergers to the AGN population are equal to, if not exceeding, the major post-mergers. Therefore, despite the highest AGN excesses occurring in the major post-mergers, the higher frequency of the mini and minor post-mergers allow them to contribute significantly to the population of the most luminous AGN. Finally, we note that even with this broader census of interactions, Figure \ref{fig:MainResult5.2} shows that in the highest luminosity bin, some 30\% of AGN have not been classified into one of our interacting categories.

Figure \ref{fig:MainResult5.2} leaves the open question of what triggers AGN in the 30\% of high luminosity cases not linked to an interaction? It is not observationally unprecedented for there to be a significant population of non-interacting AGN, in fact \cite{Genzel2014} find that AGN are relatively common in massive star forming galaxies. However, in the work presented here, we are interested in whether there are any specific characteristics common to the `non-interacting' AGN population. To answer this question, we begin by visually inspecting the stellar mass maps of galaxies that host luminous AGN but have not been identified as mergers using the methods presented thus far. We find that many of them show clear signs of tidal interactions, leading to the idea that our cuts of 500 Myr post-coalescence (for post-merger identification) and a neighbour within 100 kpc (for a pair) might not capture the full population of interacting galaxies (as expected from our own timescale results in Figure \ref{fig:MainResult4}). Indeed, given that SFRs have been shown to exhibit enhancement in pairs out to 150 kpc (e.g. \citealt{Patton2013}) and in post-mergers until 1 Gyr after coalescence \citep{Ferreira2025} there is observational precedence for extending the boundaries we have chosen. We therefore aim to establish whether merger criteria accounting for these longer enhancement timescales are more complete in capturing the luminous AGN population.

In Figure \ref{fig:MainResult5.3}, we thus plot the time since the most recent merger event with a mass ratio of at least 1:40 and the distance to the closest pair with a pair mass ratio of at least 1:40. The background density plot shows all galaxies from TNG50-1 above a stellar mass $10^{10} M_{\odot}$ and redshift less than one. The plot is separated into four quadrants based on the maximum distance (200 kpc) and time post-merger (2 Gyr) for which we see a significant AGN excess in \cite{ByrneMamahit2023} and \cite{ByrneMamahit2024}, and the occupation fractions of the total galaxy sample is shown in each quadrant. Coloured points are the galaxies from the most luminous ($L_{\text{bol}}\geq10^{45}$erg/s) AGN bin in Figure \ref{fig:MainResult5.2}. Galaxies falling within the interacting quadrants are shown in turquoise. Circles with arrows represent galaxies for which the time post-merger is a lower limit (i.e. no merger as far as the merger tree can be followed or no merger up to z=2). The remaining `non-interacting' galaxies are shown in pink. 

We find that of the 30\% of `non-interacting' galaxies hosting a luminous AGN in Figure \ref{fig:MainResult5.2}, a large number fall within the pair or post-merger limits for which we have found an excess of AGN when compared with non-merger controls. All but 4 galaxies (94\%) are accounted for within the `interaction zones' (i.e. either left of the vertical red line, or below the horizontal red line). In comparison, we find that the upper right quadrant is commonly populated in the total galaxy sample, with 40\% of all galaxies not having any merger within 2 Gyr or a neighbour within 200 kpc. We therefore find that the most luminous AGN under-populate the upper right quadrant (strictly non-interacting quadrant) when compared with the total galaxy population. In principle, if secular triggering of the most luminous AGN were common in the simulation, we could expect a comparable occupation fraction in the upper right quadrant in the most-luminous AGN as the overall population. The rarity of the most luminous AGN in the upper right quadrant suggests a strong connection between these rare events and either interactions or higher density environments. Our results are consistent with the hypothesis that the vast majority (at least 94\%, but possibly all) of these extremely luminous AGN require some type of interaction to be able to achieve the central gas densities necessary to fuel the extreme SMBH accretion rates. 

\subsection{What do mini and minor mergers look like?}
\label{subsec:whatdotheylooklike}

\begin{figure*}
    \centering
	\includegraphics[width=\textwidth]{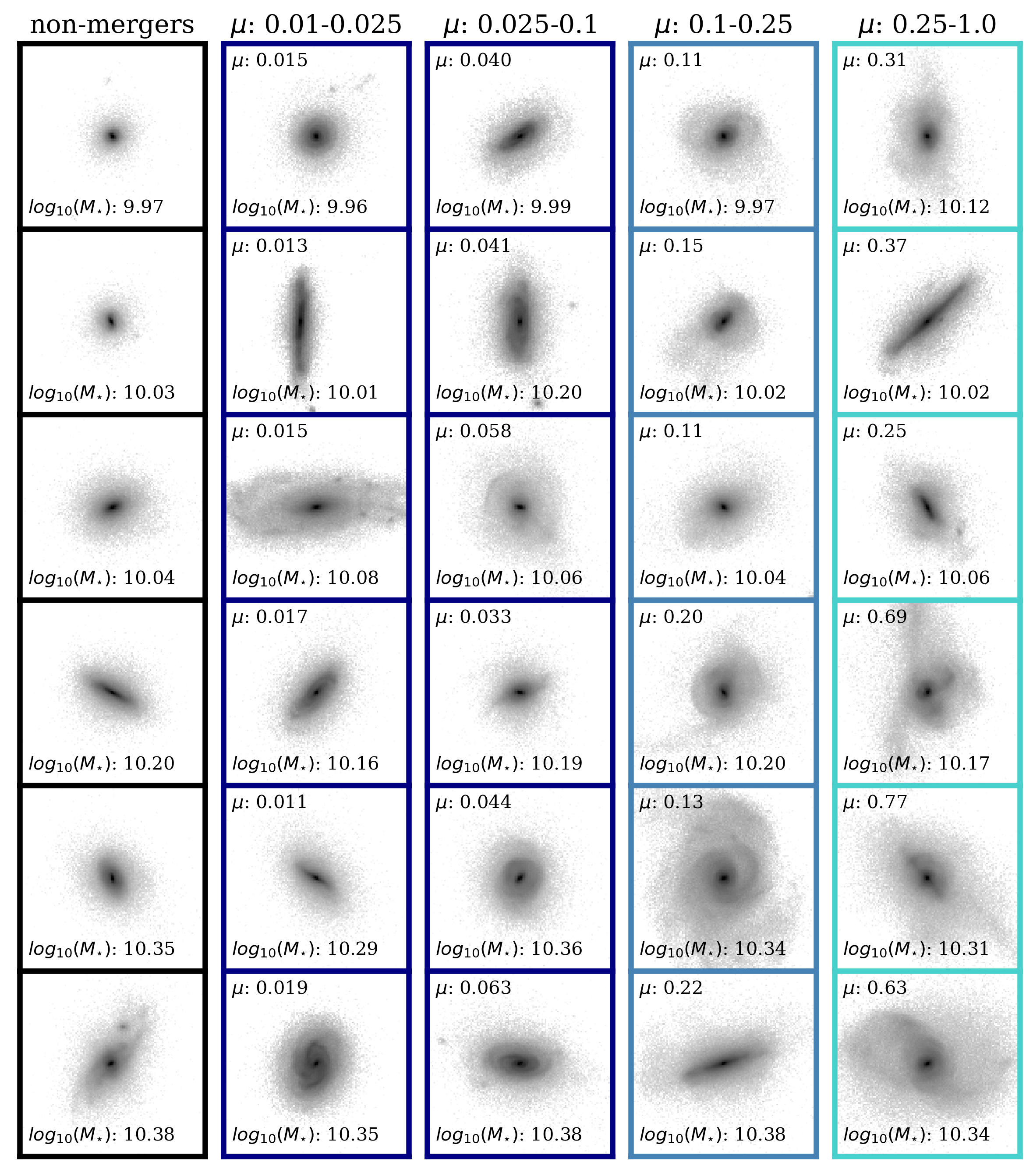}
    \caption{Stellar density maps for a selection of galaxies belonging to the categories (from left to right): non-mergers, mini post-mergers from mass ratios 1:100-1:40, mini post-mergers from mass ratios 1:40-1:10, minor post-mergers from mass ratios 1:10-1:4, and major post-mergers with mass ratios greater than 1:4. Each galaxy image is individually normalized to the highest stellar surface density pixel, and images are organized into columns by category. Galaxies in the same row are matched in mass and redshift, and the columns are organized from top to bottom in order of increasing stellar mass.}
    \label{fig:MainResult6}
\end{figure*}

In the previous sections we have demonstrated both that mini post-mergers host an excess of AGN over their matched controls, and that they contribute significantly to the populations of luminous AGN in TNG 50-1. Our results are in tension with numerous observational results that do not find the majority of AGN show visual evidence of a recent merger \citep{Schawinski2012,Hewlett2017,Marian2019}. Such observational studies rely on the presence of significant identifying features in order to classify post-mergers. However, the accurate recovery of the merger fraction using visual \citep{Lambrides2021_classification}, non-parametric \citep{Conselice2003,Wilkinson2024}, or machine vision methods \citep{Pearson2019,Bickley2021,Omori2023} is non-trivial and can vary due to numerous factors including the intrinsic properties of the interacting galaxies, the galaxy pair orbit, the time post-coalescence, and the image quality \citep{Lotz2008,Ji2014,Zeng2021,McElroy2022,DominguezSanchez2023,Bickley2024b,Wu2025}. Although there exist a number of caveats to consider when comparing simulation merger fractions to observed merger fractions, in this section we investigate the possibility that mini mergers represent a population of interaction-driven AGN that do not show obvious visual features of a recent merger.

In Figure \ref{fig:MainResult6}, we show the stellar mass density maps for a randomly selected collection of post-mergers from our TNG50-1 sample.  All of these have coalesced (according to \textsc{SubLink} merger trees) within the last $\sim$150 Myr. The vertical columns (from left to right) correspond to the following categories: galaxies that have not undergone a merger of a mass ratio greater than 1:100 within the last 2 Gyr, mini post-mergers of mass ratio 1:100-1:40, mini post-mergers of mass ratio 1:40-1:10, minor post-mergers of mass ratio 1:10-1:4, and major post-mergers of mass ratio greater than 1:4. Galaxies belonging to the same row are mass and redshift matched with each other, and galaxies are ordered by increasing mass from top to bottom. 

Broadly, we find that (as expected) the strength of visual post-merger features is strongest in the major mergers, and becomes less obvious with decreasing mass ratio. For major post-mergers, the majority of the random sample of post-mergers show visual features of being a post-merger (asymmetric morphology, shells, tidal features), although we expect that even in the major regime not all viewing angles are optimal for post-merger identification \citep{Wilkinson2024}. For post-mergers in the minor (1:10-1:4) and mini (1:40-1:10) regimes, perhaps 1/2 to 1/3 of the sample show more dramatic post-merger features. Indeed we expect some mini mergers to exhibit either obvious or subtle signatures of the interaction, as they have been shown to drive enhanced asymmetry in the galaxy population in TNG50-1 \citep{Bottrell2024}. However, the majority of the randomly selected mini post-mergers shown in Figure \ref{fig:MainResult6} galaxies do not show obvious visual indicators of being post-mergers. Similarly, post-mergers from the mass ratio regime where we do not observe a significant enhancement of SMBHAR, 1:100-1:40, are visually very consistent with the non-mergers. Qualitatively, the example stellar maps support the hypothesis that one could expect a low rate of post-merger visual recovery in the mini merger regimes. The mislabelling of these galaxies as non-interacting could explain the low fraction of post-mergers found in the most luminous AGN in observations. For our results in Figure \ref{fig:MainResult5.2}, if the mini mergers were misclassified, the post-merger fraction would be underestimated by a third.

We caution that this experiment does not rigorously determine the recovery fraction of mergers by visual identification, as it lacks radiative transfer or observational realism. However studies (e.g. \citealt{Thorp2021,Wilkinson2024,Bickley2024b}) have shown that image resolution and depth can strongly influence post-merger identification, and find the highest merger recovery fraction on the `pristine' mass maps. Therefore, observational realism would be more likely to decrease the fraction of post-mergers recovered in the luminous AGN population. We also re-iterate that our results still indicate that secular processes are responsible for the majority of the intermediate luminosity AGN population. Therefore, our results are not in tension with observational studies of low to intermediate luminosity AGN, which mostly find a non-dominant fraction of interacting galaxies. However, regardless of AGN luminosity, our results demonstrate that the fraction of mini merger induced AGN-events is likely to be significantly under-estimated in observations.

\section{Discussion}
\label{sec:discussion}

\subsection{Effects of control-matching the gas fraction in post-mergers}
\label{subsec: matching with Gas}

\begin{figure}
    \centering
	\includegraphics[width=\columnwidth]{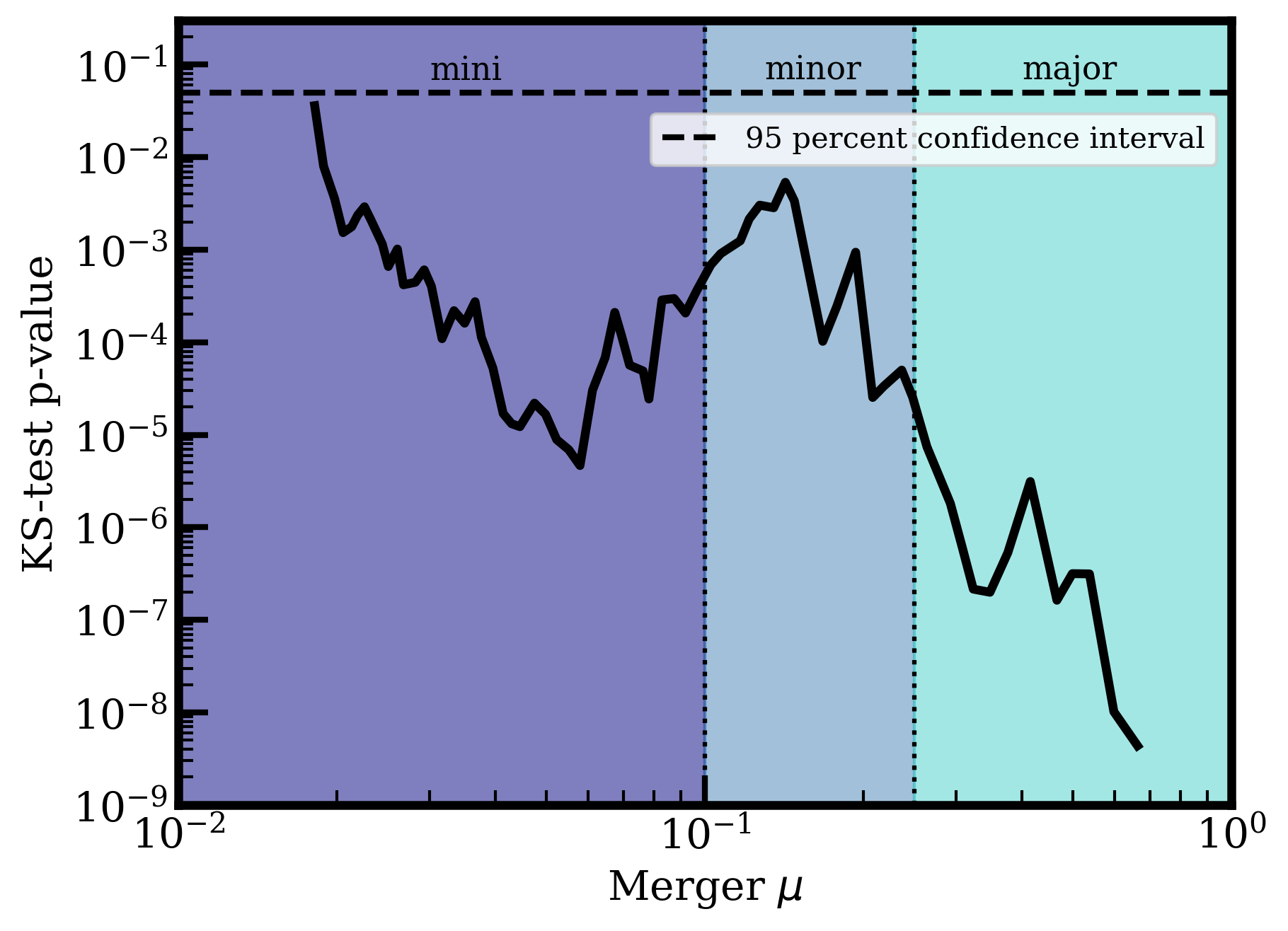}
    \caption{The probability that the distribution of the post-merger SMBH accretion rates are sampled from the same population as the matched control SMBH accretion rates, according to a KS-test, using a control-matching scheme including gas mass matching. The KS-test is performed on post-mergers (and associated controls) with a mass ratio +/- 0.1 dex from the corresponding mass ratio of the x-axis. When including gas-mass matching, the post-merger and control populations have distinguishable SMBH accretion rates for mass ratios down to 1:100.}
    \label{fig:DiscussionKSTestwithGas}
\end{figure}

\begin{figure*}
    \centering
	\includegraphics[width=\textwidth]{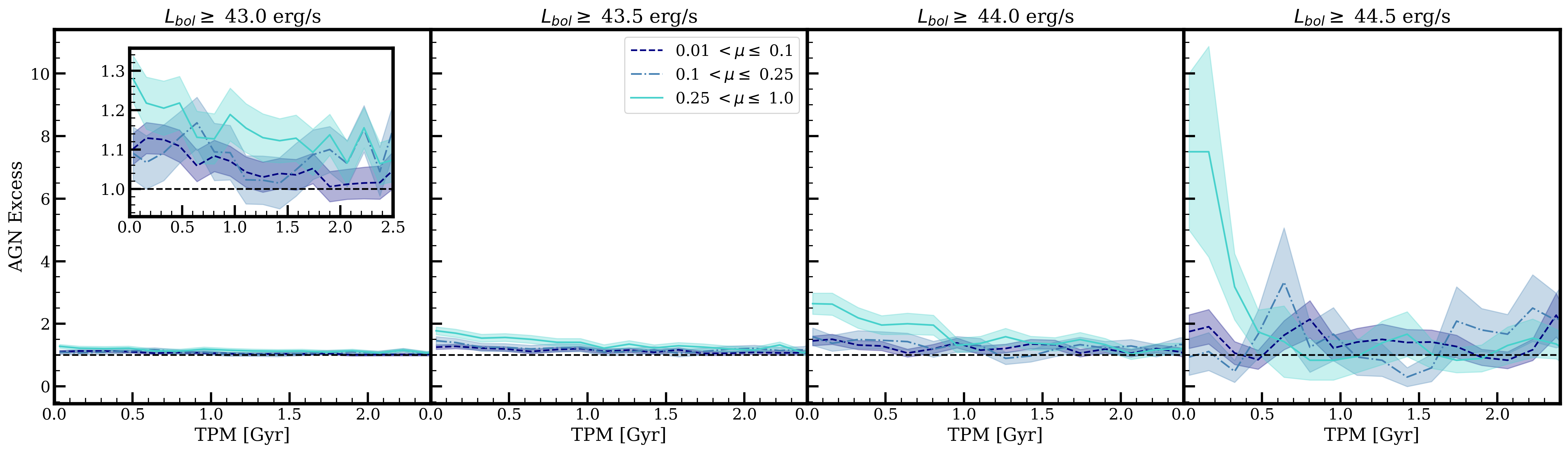}
    \caption{The AGN excess in post-mergers over the matched controls (including gas mass matching) vs the time post-merger (TPM). The mini (now defined down to 1:100-1:10), minor (1:10-1:4), and major (1:4-1:1) mergers are shown in the dark dashed, medium dash-dot, and light solid blue lines respectively. The panels are organized from left to right in order of increasing AGN luminosity threshold. The AGN excess timescale decreases with increasing AGN luminosity, and is longer in major post-mergers compared with mini and minor post-mergers.}
    \label{fig:DiscussionVsTPMwithGas}
\end{figure*}

In the analysis above we do not include gas mass in our fiducial matching scheme. The primary reason for the exclusion of gas mass matching is that it reduces the statistics of the final control-matched post-merger sample by almost 20\%. As we are investigating rare high-luminosity AGN within the post-merger population, maintaining high statistics in the final matched samples was a priority. However, the post-merger sample that we identify in TNG50-1 has a distinct distribution in gas mass from non-merger galaxies if not explicitly matched (\citealt{Schechter2025} show the enhanced gas fractions of TNG50-1 post-mergers). Given the importance of gas in fuelling AGN events, to address this point, we re-perform the analysis including gas mass matching to investigate which results are sensitive to the change in methodology.

Overall, we find nearly all of the results are qualitatively unchanged (e.g. the dependence of AGN excess on post-merger mass ratio, the long-lived AGN excess in the post-merger population), with one notable exception. In Figure \ref{fig:DiscussionKSTestwithGas} we re-perform the KS-test analysis on the SMBH accretion rates of post-merger and control galaxies, with new controls that are additionally matched on gas mass. We find that the accretion rates of post-mergers and controls are distinct across the entire mass ratio range that we explore in our analysis. We therefore find that the lower mass ratio limit for which we see a statistically significant difference in post-merger SMBHARs is dependent on the control methodology, and that mini (down to at least a mass ratio of 1:100) post-merger galaxies have SMBH accretion rates distinct from non-mergers with a similar amount of gas. It is likely that by selecting gas mass matched non-mergers, we are able to further reduce the scatter in the control matched SMBH accretion rates, and tease out a weak enhancement signal in the mini post-mergers with mass ratio between 1:100-1:40.

Figure \ref{fig:DiscussionVsTPMwithGas} shows the timescale of AGN excesses for the four AGN luminosity regimes, where we now match controls in gas mass and include post-mergers down to a mass ratio of 1:100 in the mini post-merger category. We see here that mini mergers show weaker, but statistically significant, AGN excesses for each luminosity range excepting the most luminous AGN (farthest right panel). Qualitatively, the behaviour of AGN excesses in mini, minor, and major post-mergers are unchanged: the magnitude of the AGN excess increases with increasing mass ratio, the AGN excess timescale decreases with increasing AGN luminosity and is longest in major post-mergers compared with minor and mini post-mergers.

To summarize, we find that the inclusion of gas mass matching is most significant in the determination of the minimum mass ratio over which AGN excesses are present. Including gas mass matching teases out a low level AGN excess present in mini mergers down to a mass ratio of at least 1:100.

\subsection{Effects of resolution on the dependence of AGN excess with mass ratio}
\label{subsec:resolution}

In the work presented here we have investigated the dependence of AGN triggering on mass ratios from 1:100 to 1:1 in TNG50-1. We utilized the TNG50-1 simulation as the high mass resolution allows us to select 1:100 mass ratio post-mergers while meeting our criterion of $\sim$1000 particles for the progenitor galaxies. However, in \cite{ByrneMamahit2023}, we studied SMBHAR enhancements for a sample of TNG100-1 post-mergers, down to a mass ratio of 1:10, and found that the strength of merger-induced enhancements to SMBHAR were not strongly dependent on the merger mass ratio. The results of our previous works appear to be in contrast to the work presented here, where the strength of the AGN excess is significantly dependent on the merger mass ratio. To investigate the difference here, we compare our analysis for post-mergers in TNG50-1 with both TNG100-1 (a larger volume box with a dark matter mass resolution $\sim 16$ times more massive than TNG50-1) and TNG50-2 (the lower resolution run using the same simulation initial conditions as TNG50-1, with a dark matter mass resolution $\sim8$ times more massive than TNG50-1). We restrict the analysis here to mass ratios between 1:10-1:1, as the resolution of TNG100-1 and TNG50-2 do not allow us to select 1:100-1:10 mass ratio mergers while preserving our criterion on the number of particles in the progenitor galaxies. Applying the minimum mass ratio cut of 1:10 to our TNG50-1 post-merger sample, we are left with 485 post-mergers.

We first repeat the selection and control matching of post-mergers for TNG100-1 and TNG50-2 using the exact procedure described in Section \ref{subsec:mergerID}. We find 3083 post-mergers in TNG100-1 and 405 post-mergers in TNG50-2. Performing a KS-test comparing the three post-merger samples, we find that TNG50-2 and TNG50-1 post-mergers pass a KS-test with a probability threshold of 95\% when comparing their stellar mass, environment properties ($\text{r}_1$ and $\text{N}_2$), and mass ratios. Comparing TNG100-1 and TNG50-1 post-mergers, we find that the environment and mass ratios narrowly pass a KS-test, but that the stellar masses appear to be distinct. In both TNG100-1 and TNG50-2, the star formation rate and black hole masses are distinct from TNG50-1. 

\begin{figure*}
    \centering
	\includegraphics[width=\textwidth]{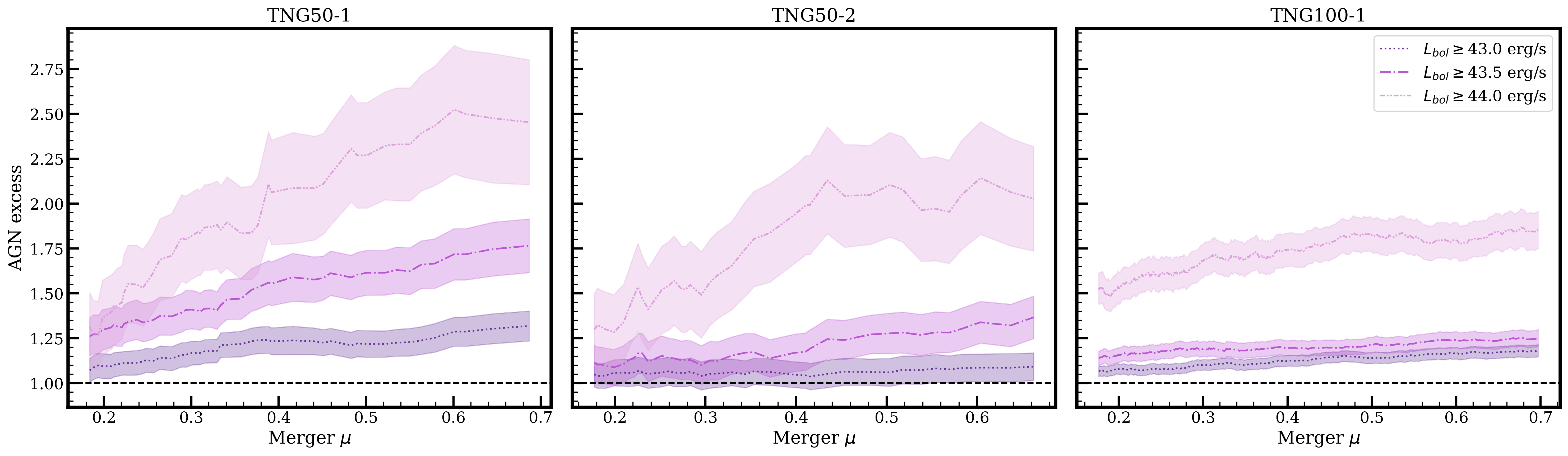}
    \caption{AGN excess (AGN fraction in post-mergers / AGN fraction in controls) as a function of merger mass ratio. The different coloured and styled lines represent the lower luminosity threshold for a galaxy to be counted as an AGN. The panels show simulations with decreasing resolution from left to right: TNG50-1 , TNG50-2, and TNG100-1.}
    \label{fig:DiscussionResComparison}
\end{figure*}

In Figure \ref{fig:DiscussionResComparison}, we repeat the experiment of calculating the AGN excess in post-mergers compared with their matched controls, as a function of the merger mass ratio for mass ratios of 1:10 to 1:1. Comparing the panels of the figure from left to right, we see that both the AGN excess magnitude and the dependence on mass ratio are weaker in lower-resolution simulations. We therefore conclude that the strong mass ratio dependence we observe in TNG50-1 that was not in our previous works is a result of resolution, and not the difference in methodology or in the initial conditions from TNG100-1 to TNG50-1.

A full analysis of the reasons behind how the change in resolution affects mass ratio dependence is beyond the scope of the work presented here, although the dependence of gas-related enhancements on simulation resolution is also found in \cite{Patton2020} who investigate SFR enhancements in paired galaxies. However, we comment that the resolution of TNG50-1 is still a significant approximation of the accretion flows expected at the levels of the accretion disk. It is expected that processes unresolved in TNG50-1 may play a significant role in the AGN-merger connection. In fact, \cite{Curtis2016} find that sufficient resolution to capture the high angular momentum of nuclear gas plays a role in delaying the transport of gas to the SMBH, delaying the onset of quasar activity in idealized disk mergers. Furthermore, the high resolution simulations in \cite{Angles2021} demonstrate the complex geometries expected in the nuclear region with multiple misaligned accretion disks and gaps within 10-100 pc scales (recalling that the softening length of gas cells in TNG50-1 is 74 pc).

Bridging the sub-parsec scale physics governing accretion flows onto the SMBH to the megaparsec scale cosmological evolution of the Universe is a significant challenge, and requires continued research into the processes unresolved in cosmological simulations, such as the dynamic evolution of the accretion disk geometry over time \citep{Angles2021}, or the evolution of SMBH binaries \citep{Cuadra2009, Liao2024}. Luckily, there are efforts to improve sub-grid implementations in cosmological simulations, which will hopefully allow for future works to further investigate how these complex unresolved processes affect AGN in post-mergers \citep{Koudmani2024,Weinberger2025}.

\section{Conclusions}
\label{sec:conclusion}

In the work presented here, we study post-merger galaxies from the Illustris TNG50-1 simulation, investigating AGN triggering to an unprecedented lower mass ratio limit of 1:100. Our sample consists of 1179 successfully control matched post-mergers, which are compared with 5895 non-merger control galaxies (5 control galaxies per post-merger). Our results are summarized in the following points.

\begin{itemize}
    \item \textbf{The SMBH accretion rates of the post-mergers are statistically distinguishable from their controls down to a minimum mass ratio of 1:40.} In Figure \ref{fig:MainResult2} we demonstrate that the SMBHAR distribution of post-mergers is statistically distinguishable from non-merger controls down to a mass ratio of 1:40. However, we further note in Section \ref{subsec: matching with Gas} that if we compare the post-mergers with non-merger controls matched in gas mass that the accretion rates are distinguishable down to a mass ratio of at least 1:100.
    \medskip
    \item \textbf{The most luminous AGN are only present in excess in major mergers, $\mu \geq$ 1:4.} In Figure \ref{fig:MainResult3}, we find that an excess of intermediate luminosity AGN, $L_{\text{bol}} \geq 10^{43-44}$ erg/s, is present down to a mass ratio of 1:40, but that the most luminous AGN, $L_{\text{bol}} \geq 10^{44.5}$ erg/s, only demonstrate a statistically significant AGN excess above mass ratios of 1:4.
    \medskip
    \item \textbf{The magnitude of the AGN excess is luminosity dependent.} In Figure \ref{fig:MainResult3}, for intermediate luminosity AGN, the maximum AGN excess in post-mergers is between 2-3 times higher than non-merger controls. For the most luminous AGN, the maximum excess is approximately 7.
    \medskip
    \item \textbf{Mini (1:40$\leq \mu <$1:10), minor (1:10$\leq \mu <$1:4), and major (1:4$ \leq \mu <$1:1) mergers exhibit long-lived AGN excesses, and the timescale of the AGN excess is luminosity and mass ratio dependent, decreasing with increasing luminosity and longest in major mergers.} In Figure \ref{fig:MainResult4}, we find AGN excesses on timescales up to 1-2 Gyr in low(er) luminosity AGN, and 500 Myr for major mergers in the highest luminosity AGN.
    \medskip
    \item \textbf{The majority (>50\%) of the most luminous AGN are found in interacting systems.} In Figure \ref{fig:MainResult5.2}, the fraction of the AGN in mini + minor mergers equals or exceeds the contribution from the major mergers. Additionally, in Figure \ref{fig:MainResult5.3} we find that 94 \% of the most luminous ($L_{\text{bol}}\geq 10^{45}$erg/s) AGN are found within the post-merger or pair regime for which we see any statistical enhancement of AGN (i.e. either: a post-merger within 2 Gyr or a pair within 200 kpc). 
    \medskip
    \item \textbf{Mini merger fractions are likely to be underestimated in observations.} Using a qualitative visual inspection of a random selection of mini post-mergers, in Figure \ref{fig:MainResult6} we find that the majority of mini post-mergers do not exhibit noticeable post-merger features, and may therefore constitute a `hidden' merger population that could be mis-labelled as non-interacting.
\end{itemize}

Overall, our results demonstrate that a complete investigation into the role of interactions in galaxy evolution must also include mini and minor mergers. In the work here, we investigate the population of mini mergers, and how the AGN excess in the population varies with mass ratio. However, we do not investigate the specific details of how the gaseous inflows responsible for triggering the AGN may differ between the different mass ratio regimes. A more detailed investigation could provide further insights into the different inflow mechanisms and their consequences for the characteristics of the post-merger galaxies.

\section*{Acknowledgements}
We acknowledge and thank the IllustrisTNG collaboration for providing public access to data from the TNG simulations. SJB acknowledges the receipt of the Dr. Margaret Perkins Hess Research Fellowship from the University of Victoria and SJB and SW acknowledge graduate fellowships from the Natural Sciences and Engineering Research Council of Canada (NSERC);  Cette recherche a été financée par le Conseil de recherches en sciences naturelles et en génie du Canada (CRSNG). This research was enabled in part by support provided by WestGrid (www.westgrid.ca) and the Digital Research Alliance of Canada (alliancecan.ca). SLE and DRP gratefully acknowledge the NSERC of Canada for Discovery Grants which helped to fund this research.

\section*{Data Availability}
The data used in this work are publicly available at \hyperlink{}{https://www.tng-project.org}.



\bibliographystyle{mnras}
\bibliography{paperFinal}

\begin{thebibliography}{}
\makeatletter
\relax
\def\mn@urlcharsother{\let\do\@makeother \do\$\do\&\do\#\do\^\do\_\do\%\do\~}
\def\mn@doi{\begingroup\mn@urlcharsother \@ifnextchar [ {\mn@doi@}
  {\mn@doi@[]}}
\def\mn@doi@[#1]#2{\def\@tempa{#1}\ifx\@tempa\@empty \href
  {http://dx.doi.org/#2} {doi:#2}\else \href {http://dx.doi.org/#2} {#1}\fi
  \endgroup}
\def\mn@eprint#1#2{\mn@eprint@#1:#2::\@nil}
\def\mn@eprint@arXiv#1{\href {http://arxiv.org/abs/#1} {{\tt arXiv:#1}}}
\def\mn@eprint@dblp#1{\href {http://dblp.uni-trier.de/rec/bibtex/#1.xml}
  {dblp:#1}}
\def\mn@eprint@#1:#2:#3:#4\@nil{\def\@tempa {#1}\def\@tempb {#2}\def\@tempc
  {#3}\ifx \@tempc \@empty \let \@tempc \@tempb \let \@tempb \@tempa \fi \ifx
  \@tempb \@empty \def\@tempb {arXiv}\fi \@ifundefined
  {mn@eprint@\@tempb}{\@tempb:\@tempc}{\expandafter \expandafter \csname
  mn@eprint@\@tempb\endcsname \expandafter{\@tempc}}}

\bibitem[\protect\citeauthoryear{{Alonso}, {Lambas}, {Tissera}  \&
  {Coldwell}}{{Alonso} et~al.}{2007}]{Alonso2007}
{Alonso} M.~S.,  {Lambas} D.~G.,  {Tissera} P.,   {Coldwell} G.,  2007, \mn@doi
  [\mnras] {10.1111/j.1365-2966.2007.11367.x}, \href
  {https://ui.adsabs.harvard.edu/abs/2007MNRAS.375.1017A} {375, 1017}

\bibitem[\protect\citeauthoryear{{Angl{\'e}s-Alc{\'a}zar}
  et~al.,}{{Angl{\'e}s-Alc{\'a}zar} et~al.}{2021}]{Angles2021}
{Angl{\'e}s-Alc{\'a}zar} D.,  et~al., 2021, \mn@doi [\apj]
  {10.3847/1538-4357/ac09e8}, \href
  {https://ui.adsabs.harvard.edu/abs/2021ApJ...917...53A} {917, 53}

\bibitem[\protect\citeauthoryear{{Bah{\'e}} et~al.,}{{Bah{\'e}}
  et~al.}{2022}]{Bahe2022}
{Bah{\'e}} Y.~M.,  et~al., 2022, \mn@doi [\mnras] {10.1093/mnras/stac1339},
  \href {https://ui.adsabs.harvard.edu/abs/2022MNRAS.516..167B} {516, 167}

\bibitem[\protect\citeauthoryear{{Barnes} \& {Hernquist}}{{Barnes} \&
  {Hernquist}}{1991}]{Barnes1991}
{Barnes} J.~E.,  {Hernquist} L.~E.,  1991, \mn@doi [\apjl] {10.1086/185978},
  \href {https://ui.adsabs.harvard.edu/abs/1991ApJ...370L..65B} {370, L65}

\bibitem[\protect\citeauthoryear{{Barton}, {Geller}  \& {Kenyon}}{{Barton}
  et~al.}{2000}]{Barton2000}
{Barton} E.~J.,  {Geller} M.~J.,   {Kenyon} S.~J.,  2000, \mn@doi [\apj]
  {10.1086/308392}, \href
  {https://ui.adsabs.harvard.edu/abs/2000ApJ...530..660B} {530, 660}

\bibitem[\protect\citeauthoryear{{Bennert}, {Canalizo}, {Jungwiert},
  {Stockton}, {Schweizer}, {Peng}  \& {Lacy}}{{Bennert}
  et~al.}{2008}]{Bennert2008}
{Bennert} N.,  {Canalizo} G.,  {Jungwiert} B.,  {Stockton} A.,  {Schweizer} F.,
   {Peng} C.,   {Lacy} M.,  2008, \memsai, \href
  {https://ui.adsabs.harvard.edu/abs/2008MmSAI..79.1247B} {79, 1247}

\bibitem[\protect\citeauthoryear{{Bessiere}, {Tadhunter}, {Ramos Almeida}  \&
  {Villar Mart{\'\i}n}}{{Bessiere} et~al.}{2012}]{Bessiere2012}
{Bessiere} P.~S.,  {Tadhunter} C.~N.,  {Ramos Almeida} C.,   {Villar
  Mart{\'\i}n} M.,  2012, \mn@doi [\mnras] {10.1111/j.1365-2966.2012.21701.x},
  \href {https://ui.adsabs.harvard.edu/abs/2012MNRAS.426..276B} {426, 276}

\bibitem[\protect\citeauthoryear{{Bickley} et~al.,}{{Bickley}
  et~al.}{2021}]{Bickley2021}
{Bickley} R.~W.,  et~al., 2021, \mn@doi [\mnras] {10.1093/mnras/stab806}, \href
  {https://ui.adsabs.harvard.edu/abs/2021MNRAS.504..372B} {504, 372}

\bibitem[\protect\citeauthoryear{{Bickley}, {Ellison}, {Patton}, {Bottrell},
  {Gwyn}  \& {Hudson}}{{Bickley} et~al.}{2022}]{Bickley2022}
{Bickley} R.~W.,  {Ellison} S.~L.,  {Patton} D.~R.,  {Bottrell} C.,  {Gwyn} S.,
    {Hudson} M.~J.,  2022, \mn@doi [\mnras] {10.1093/mnras/stac1500}, \href
  {https://ui.adsabs.harvard.edu/abs/2022MNRAS.tmp.1474B} {}

\bibitem[\protect\citeauthoryear{{Bickley}, {Ellison}, {Patton}  \&
  {Wilkinson}}{{Bickley} et~al.}{2023}]{Bickley2023}
{Bickley} R.~W.,  {Ellison} S.~L.,  {Patton} D.~R.,   {Wilkinson} S.,  2023,
  \mn@doi [\mnras] {10.1093/mnras/stad088}, \href
  {https://ui.adsabs.harvard.edu/abs/2023MNRAS.519.6149B} {519, 6149}

\bibitem[\protect\citeauthoryear{{Bickley} et~al.,}{{Bickley}
  et~al.}{2024a}]{Bickley2024a}
{Bickley} R.~W.,  et~al., 2024a, \mn@doi [\mnras] {10.1093/mnras/stae1951},
  \href {https://ui.adsabs.harvard.edu/abs/2024MNRAS.533.3068B} {533, 3068}

\bibitem[\protect\citeauthoryear{{Bickley}, {Wilkinson}, {Ferreira}, {Ellison},
  {Bottrell}  \& {Jyoti}}{{Bickley} et~al.}{2024b}]{Bickley2024b}
{Bickley} R.~W.,  {Wilkinson} S.,  {Ferreira} L.,  {Ellison} S.~L.,  {Bottrell}
  C.,   {Jyoti} D.,  2024b, \mn@doi [\mnras] {10.1093/mnras/stae2246}, \href
  {https://ui.adsabs.harvard.edu/abs/2024MNRAS.534.2533B} {534, 2533}

\bibitem[\protect\citeauthoryear{{Blecha}, {Snyder}, {Satyapal}  \&
  {Ellison}}{{Blecha} et~al.}{2018}]{Blecha2018}
{Blecha} L.,  {Snyder} G.~F.,  {Satyapal} S.,   {Ellison} S.~L.,  2018, \mn@doi
  [\mnras] {10.1093/mnras/sty1274}, \href
  {https://ui.adsabs.harvard.edu/abs/2018MNRAS.478.3056B} {478, 3056}

\bibitem[\protect\citeauthoryear{Blumenthal \& Barnes}{Blumenthal \&
  Barnes}{2018}]{Blumenthal2018}
Blumenthal K.~A.,  Barnes J.~E.,  2018, \mn@doi [\mnras]
  {10.1093/mnras/sty1605}, 479, 3952

\bibitem[\protect\citeauthoryear{{Blumenthal}, {Faber}, {Primack}  \&
  {Rees}}{{Blumenthal} et~al.}{1984}]{Blumenthal1984}
{Blumenthal} G.~R.,  {Faber} S.~M.,  {Primack} J.~R.,   {Rees} M.~J.,  1984,
  \mn@doi [\nat] {10.1038/311517a0}, \href
  {https://ui.adsabs.harvard.edu/abs/1984Natur.311..517B} {311, 517}

\bibitem[\protect\citeauthoryear{{B{\"o}hm} et~al.,}{{B{\"o}hm}
  et~al.}{2013}]{Bohm2013}
{B{\"o}hm} A.,  et~al., 2013, \mn@doi [\aap] {10.1051/0004-6361/201015444},
  \href {https://ui.adsabs.harvard.edu/abs/2013A&A...549A..46B} {549, A46}

\bibitem[\protect\citeauthoryear{{Bondi}}{{Bondi}}{1952}]{Bondi1952}
{Bondi} H.,  1952, \mn@doi [\mnras] {10.1093/mnras/112.2.195}, \href
  {https://ui.adsabs.harvard.edu/abs/1952MNRAS.112..195B} {112, 195}

\bibitem[\protect\citeauthoryear{{Bondi} \& {Hoyle}}{{Bondi} \&
  {Hoyle}}{1944}]{Bondi1944}
{Bondi} H.,  {Hoyle} F.,  1944, \mn@doi [\mnras] {10.1093/mnras/104.5.273},
  \href {https://ui.adsabs.harvard.edu/abs/1944MNRAS.104..273B} {104, 273}

\bibitem[\protect\citeauthoryear{{Bottrell} et~al.,}{{Bottrell}
  et~al.}{2024}]{Bottrell2024}
{Bottrell} C.,  et~al., 2024, \mn@doi [\mnras] {10.1093/mnras/stad2971}, \href
  {https://ui.adsabs.harvard.edu/abs/2024MNRAS.527.6506B} {527, 6506}

\bibitem[\protect\citeauthoryear{{Buttigieg}, {Sijacki}, {Moore}  \&
  {Bourne}}{{Buttigieg} et~al.}{2025}]{Buttigieg2025}
{Buttigieg} S.,  {Sijacki} D.,  {Moore} C.~J.,   {Bourne} M.~A.,  2025, \mn@doi
  [arXiv e-prints] {10.48550/arXiv.2504.17549}, \href
  {https://ui.adsabs.harvard.edu/abs/2025arXiv250417549B} {p. arXiv:2504.17549}

\bibitem[\protect\citeauthoryear{{Byrne-Mamahit}, {Hani}, {Ellison}, {Quai}  \&
  {Patton}}{{Byrne-Mamahit} et~al.}{2023}]{ByrneMamahit2023}
{Byrne-Mamahit} S.,  {Hani} M.~H.,  {Ellison} S.~L.,  {Quai} S.,   {Patton}
  D.~R.,  2023, \mn@doi [\mnras] {10.1093/mnras/stac3674}, \href
  {https://ui.adsabs.harvard.edu/abs/2023MNRAS.519.4966B} {519, 4966}

\bibitem[\protect\citeauthoryear{{Byrne-Mamahit}, {Patton}, {Ellison},
  {Bickley}, {Ferreira}, {Hani}, {Quai}  \& {Wilkinson}}{{Byrne-Mamahit}
  et~al.}{2024}]{ByrneMamahit2024}
{Byrne-Mamahit} S.,  {Patton} D.~R.,  {Ellison} S.~L.,  {Bickley} R.,
  {Ferreira} L.,  {Hani} M.,  {Quai} S.,   {Wilkinson} S.,  2024, \mn@doi
  [\mnras] {10.1093/mnras/stae419}, \href
  {https://ui.adsabs.harvard.edu/abs/2024MNRAS.528.5864B} {528, 5864}

\bibitem[\protect\citeauthoryear{{Cao} et~al.,}{{Cao} et~al.}{2016}]{Cao2016}
{Cao} C.,  et~al., 2016, \mn@doi [\apjs] {10.3847/0067-0049/222/2/16}, \href
  {https://ui.adsabs.harvard.edu/abs/2016ApJS..222...16C} {222, 16}

\bibitem[\protect\citeauthoryear{{Capelo} \& {Dotti}}{{Capelo} \&
  {Dotti}}{2017}]{CapeloDotti2017}
{Capelo} P.~R.,  {Dotti} M.,  2017, \mn@doi [\mnras] {10.1093/mnras/stw2872},
  \href {https://ui.adsabs.harvard.edu/abs/2017MNRAS.465.2643C} {465, 2643}

\bibitem[\protect\citeauthoryear{{Capelo}, {Volonteri}, {Dotti}, {Bellovary},
  {Mayer}  \& {Governato}}{{Capelo} et~al.}{2015}]{Capelo2015}
{Capelo} P.~R.,  {Volonteri} M.,  {Dotti} M.,  {Bellovary} J.~M.,  {Mayer} L.,
   {Governato} F.,  2015, \mn@doi [\mnras] {10.1093/mnras/stu2500}, \href
  {https://ui.adsabs.harvard.edu/abs/2015MNRAS.447.2123C} {447, 2123}

\bibitem[\protect\citeauthoryear{{Choi}, {Somerville}, {Ostriker}, {Hirschmann}
   \& {Naab}}{{Choi} et~al.}{2024}]{Choi2024}
{Choi} E.,  {Somerville} R.~S.,  {Ostriker} J.~P.,  {Hirschmann} M.,   {Naab}
  T.,  2024, \mn@doi [\apj] {10.3847/1538-4357/ad245a}, \href
  {https://ui.adsabs.harvard.edu/abs/2024ApJ...964...54C} {964, 54}

\bibitem[\protect\citeauthoryear{{Cisternas} et~al.,}{{Cisternas}
  et~al.}{2011}]{Cisternas2011}
{Cisternas} M.,  et~al., 2011, \mn@doi [\apj] {10.1088/0004-637X/726/2/57},
  \href {https://ui.adsabs.harvard.edu/abs/2011ApJ...726...57C} {726, 57}

\bibitem[\protect\citeauthoryear{{Comerford} et~al.,}{{Comerford}
  et~al.}{2024}]{Comerford2024}
{Comerford} J.~M.,  et~al., 2024, \mn@doi [\apj] {10.3847/1538-4357/ad1a15},
  \href {https://ui.adsabs.harvard.edu/abs/2024ApJ...963...53C} {963, 53}

\bibitem[\protect\citeauthoryear{{Conselice}, {Bershady}, {Dickinson}  \&
  {Papovich}}{{Conselice} et~al.}{2003}]{Conselice2003}
{Conselice} C.~J.,  {Bershady} M.~A.,  {Dickinson} M.,   {Papovich} C.,  2003,
  \mn@doi [\aj] {10.1086/377318}, \href
  {https://ui.adsabs.harvard.edu/abs/2003AJ....126.1183C} {126, 1183}

\bibitem[\protect\citeauthoryear{{Conselice}, {Mundy}, {Ferreira}  \&
  {Duncan}}{{Conselice} et~al.}{2022}]{Conselice2022}
{Conselice} C.~J.,  {Mundy} C.~J.,  {Ferreira} L.,   {Duncan} K.,  2022,
  \mn@doi [\apj] {10.3847/1538-4357/ac9b1a}, \href
  {https://ui.adsabs.harvard.edu/abs/2022ApJ...940..168C} {940, 168}

\bibitem[\protect\citeauthoryear{{Cox}, {Jonsson}, {Somerville}, {Primack}  \&
  {Dekel}}{{Cox} et~al.}{2008}]{Cox2008}
{Cox} T.~J.,  {Jonsson} P.,  {Somerville} R.~S.,  {Primack} J.~R.,   {Dekel}
  A.,  2008, \mn@doi [\mnras] {10.1111/j.1365-2966.2007.12730.x}, \href
  {https://ui.adsabs.harvard.edu/abs/2008MNRAS.384..386C} {384, 386}

\bibitem[\protect\citeauthoryear{{Cuadra}, {Armitage}, {Alexander}  \&
  {Begelman}}{{Cuadra} et~al.}{2009}]{Cuadra2009}
{Cuadra} J.,  {Armitage} P.~J.,  {Alexander} R.~D.,   {Begelman} M.~C.,  2009,
  \mn@doi [\mnras] {10.1111/j.1365-2966.2008.14147.x}, \href
  {https://ui.adsabs.harvard.edu/abs/2009MNRAS.393.1423C} {393, 1423}

\bibitem[\protect\citeauthoryear{{Curtis} \& {Sijacki}}{{Curtis} \&
  {Sijacki}}{2016}]{Curtis2016}
{Curtis} M.,  {Sijacki} D.,  2016, \mn@doi [\mnras] {10.1093/mnras/stw1944},
  \href {https://ui.adsabs.harvard.edu/abs/2016MNRAS.463...63C} {463, 63}

\bibitem[\protect\citeauthoryear{{Davies}, {Crain}, {Oppenheimer}  \&
  {Schaye}}{{Davies} et~al.}{2020}]{Davies2020}
{Davies} J.~J.,  {Crain} R.~A.,  {Oppenheimer} B.~D.,   {Schaye} J.,  2020,
  \mn@doi [\mnras] {10.1093/mnras/stz3201}, \href
  {https://ui.adsabs.harvard.edu/abs/2020MNRAS.491.4462D} {491, 4462}

\bibitem[\protect\citeauthoryear{{Di Matteo}, {Springel}  \& {Hernquist}}{{Di
  Matteo} et~al.}{2005}]{DiMatteo2005}
{Di Matteo} T.,  {Springel} V.,   {Hernquist} L.,  2005, \mn@doi [\nat]
  {10.1038/nature03335}, \href
  {https://ui.adsabs.harvard.edu/abs/2005Natur.433..604D} {433, 604}

\bibitem[\protect\citeauthoryear{{Di Matteo}, {Combes}, {Melchior}  \&
  {Semelin}}{{Di Matteo} et~al.}{2007}]{DiMatteo2007}
{Di Matteo} P.,  {Combes} F.,  {Melchior} A.~L.,   {Semelin} B.,  2007, \mn@doi
  [\aap] {10.1051/0004-6361:20066959}, \href
  {https://ui.adsabs.harvard.edu/abs/2007A&A...468...61D} {468, 61}

\bibitem[\protect\citeauthoryear{{Dom{\'\i}nguez S{\'a}nchez}
  et~al.,}{{Dom{\'\i}nguez S{\'a}nchez} et~al.}{2023}]{DominguezSanchez2023}
{Dom{\'\i}nguez S{\'a}nchez} H.,  et~al., 2023, \mn@doi [\mnras]
  {10.1093/mnras/stad750}, \href
  {https://ui.adsabs.harvard.edu/abs/2023MNRAS.521.3861D} {521, 3861}

\bibitem[\protect\citeauthoryear{{Dougherty}, {Harrison}, {Kocevski}  \&
  {Rosario}}{{Dougherty} et~al.}{2024}]{Dougherty2024}
{Dougherty} S.~L.,  {Harrison} C.~M.,  {Kocevski} D.~D.,   {Rosario} D.~J.,
  2024, \mn@doi [\mnras] {10.1093/mnras/stad1300}, \href
  {https://ui.adsabs.harvard.edu/abs/2024MNRAS.527.3146D} {527, 3146}

\bibitem[\protect\citeauthoryear{{Dunn}, {Fender}, {K{\"o}rding}, {Belloni}  \&
  {Cabanac}}{{Dunn} et~al.}{2010}]{Dunn2010}
{Dunn} R.~J.~H.,  {Fender} R.~P.,  {K{\"o}rding} E.~G.,  {Belloni} T.,
  {Cabanac} C.,  2010, \mn@doi [\mnras] {10.1111/j.1365-2966.2010.16114.x},
  \href {https://ui.adsabs.harvard.edu/abs/2010MNRAS.403...61D} {403, 61}

\bibitem[\protect\citeauthoryear{Ellison, Patton, Simard  \&
  McConnachie}{Ellison et~al.}{2008}]{Ellison2008}
Ellison S.~L.,  Patton D.~R.,  Simard L.,   McConnachie A.~W.,  2008, \mn@doi
  [\aj] {10.1088/0004-6256/135/5/1877}, 135, 1877

\bibitem[\protect\citeauthoryear{{Ellison}, {Patton}, {Mendel}  \&
  {Scudder}}{{Ellison} et~al.}{2011}]{Ellison2011}
{Ellison} S.~L.,  {Patton} D.~R.,  {Mendel} J.~T.,   {Scudder} J.~M.,  2011,
  \mn@doi [\mnras] {10.1111/j.1365-2966.2011.19624.x}, \href
  {https://ui.adsabs.harvard.edu/abs/2011MNRAS.418.2043E} {418, 2043}

\bibitem[\protect\citeauthoryear{{Ellison}, {Mendel}, {Patton}  \&
  {Scudder}}{{Ellison} et~al.}{2013}]{Ellison2013}
{Ellison} S.~L.,  {Mendel} J.~T.,  {Patton} D.~R.,   {Scudder} J.~M.,  2013,
  \mn@doi [\mnras] {10.1093/mnras/stt1562}, \href
  {https://ui.adsabs.harvard.edu/abs/2013MNRAS.435.3627E} {435, 3627}

\bibitem[\protect\citeauthoryear{{Ellison} et~al.,}{{Ellison}
  et~al.}{2022}]{Ellison2022}
{Ellison} S.~L.,  et~al., 2022, \mn@doi [\mnras] {10.1093/mnrasl/slac109},
  \href {https://ui.adsabs.harvard.edu/abs/2022MNRAS.517L..92E} {517, L92}

\bibitem[\protect\citeauthoryear{{Ellison}, {Ferreira}, {Wild}, {Wilkinson},
  {Rowlands}  \& {Patton}}{{Ellison} et~al.}{2024}]{Ellison2024}
{Ellison} S.,  {Ferreira} L.,  {Wild} V.,  {Wilkinson} S.,  {Rowlands} K.,
  {Patton} D.~R.,  2024, \mn@doi [The Open Journal of Astrophysics]
  {10.33232/001c.127779}, \href
  {https://ui.adsabs.harvard.edu/abs/2024OJAp....7E.121E} {7, 121}

\bibitem[\protect\citeauthoryear{{Ellison} et~al.,}{{Ellison}
  et~al.}{2025}]{Ellison2025}
{Ellison} S.,  et~al., 2025, \mn@doi [The Open Journal of Astrophysics]
  {10.33232/001c.129235}, \href
  {https://ui.adsabs.harvard.edu/abs/2025OJAp....8E..12E} {8, 12}

\bibitem[\protect\citeauthoryear{{Faria}, {Patton}, {Courteau}, {Ellison}  \&
  {Brown}}{{Faria} et~al.}{2025}]{Faria2025}
{Faria} L.,  {Patton} D.~R.,  {Courteau} S.,  {Ellison} S.,   {Brown} W.,
  2025, \mn@doi [\mnras] {10.1093/mnras/staf124}, \href
  {https://ui.adsabs.harvard.edu/abs/2025MNRAS.537..915F} {537, 915}

\bibitem[\protect\citeauthoryear{{Ferreira} et~al.,}{{Ferreira}
  et~al.}{2024}]{Ferreira2024}
{Ferreira} L.,  et~al., 2024, \mn@doi [\mnras] {10.1093/mnras/stae1885}, \href
  {https://ui.adsabs.harvard.edu/abs/2024MNRAS.533.2547F} {533, 2547}

\bibitem[\protect\citeauthoryear{{Ferreira}, {Ellison}, {Patton},
  {Byrne-Mamahit}, {Wilkinson}, {Bickley}, {Conselice}  \&
  {Bottrell}}{{Ferreira} et~al.}{2025}]{Ferreira2025}
{Ferreira} L.,  {Ellison} S.~L.,  {Patton} D.~R.,  {Byrne-Mamahit} S.,
  {Wilkinson} S.,  {Bickley} R.,  {Conselice} C.~J.,   {Bottrell} C.,  2025,
  \mn@doi [\mnras] {10.1093/mnrasl/slaf004}, \href
  {https://ui.adsabs.harvard.edu/abs/2025MNRAS.538L..31F} {538, L31}

\bibitem[\protect\citeauthoryear{{Genzel} et~al.,}{{Genzel}
  et~al.}{2014}]{Genzel2014}
{Genzel} R.,  et~al., 2014, \mn@doi [\apj] {10.1088/0004-637X/796/1/7}, \href
  {https://ui.adsabs.harvard.edu/abs/2014ApJ...796....7G} {796, 7}

\bibitem[\protect\citeauthoryear{{Glikman}, {Simmons}, {Mailly}, {Schawinski},
  {Urry}  \& {Lacy}}{{Glikman} et~al.}{2015}]{Glikman2015}
{Glikman} E.,  {Simmons} B.,  {Mailly} M.,  {Schawinski} K.,  {Urry} C.~M.,
  {Lacy} M.,  2015, \mn@doi [\apj] {10.1088/0004-637X/806/2/218}, \href
  {https://ui.adsabs.harvard.edu/abs/2015ApJ...806..218G} {806, 218}

\bibitem[\protect\citeauthoryear{Hani, Gosain, Ellison, Patton  \& Torrey}{Hani
  et~al.}{2020}]{Hani2020}
Hani M.~H.,  Gosain H.,  Ellison S.~L.,  Patton D.~R.,   Torrey P.,  2020,
  \mn@doi [\mnras] {10.1093/mnras/staa459}, 493, 3716

\bibitem[\protect\citeauthoryear{{Hernquist}}{{Hernquist}}{1989}]{Hernquist1989a}
{Hernquist} L.,  1989, \mn@doi [\nat] {10.1038/340687a0}, \href
  {https://ui.adsabs.harvard.edu/abs/1989Natur.340..687H} {340, 687}

\bibitem[\protect\citeauthoryear{{Hewlett}, {Villforth}, {Wild},
  {Mendez-Abreu}, {Pawlik}  \& {Rowlands}}{{Hewlett}
  et~al.}{2017}]{Hewlett2017}
{Hewlett} T.,  {Villforth} C.,  {Wild} V.,  {Mendez-Abreu} J.,  {Pawlik} M.,
  {Rowlands} K.,  2017, \mn@doi [\mnras] {10.1093/mnras/stx997}, \href
  {https://ui.adsabs.harvard.edu/abs/2017MNRAS.470..755H} {470, 755}

\bibitem[\protect\citeauthoryear{Hong, Im, Kim  \& Ho}{Hong
  et~al.}{2015}]{Hong2015}
Hong J.,  Im M.,  Kim M.,   Ho L.~C.,  2015, \mn@doi [\apj]
  {10.1088/0004-637x/804/1/34}, 804, 34

\bibitem[\protect\citeauthoryear{{Hopkins}, {Hernquist}, {Cox}  \&
  {Kere{\v{s}}}}{{Hopkins} et~al.}{2008}]{Hopkins2008}
{Hopkins} P.~F.,  {Hernquist} L.,  {Cox} T.~J.,   {Kere{\v{s}}} D.,  2008,
  \mn@doi [\apjs] {10.1086/524362}, \href
  {https://ui.adsabs.harvard.edu/abs/2008ApJS..175..356H} {175, 356}

\bibitem[\protect\citeauthoryear{{Ji}, {Peirani}  \& {Yi}}{{Ji}
  et~al.}{2014}]{Ji2014}
{Ji} I.,  {Peirani} S.,   {Yi} S.~K.,  2014, \mn@doi [\aap]
  {10.1051/0004-6361/201423530}, \href
  {https://ui.adsabs.harvard.edu/abs/2014A&A...566A..97J} {566, A97}

\bibitem[\protect\citeauthoryear{{Johansson}, {Naab}  \& {Burkert}}{{Johansson}
  et~al.}{2009}]{Johansson2009}
{Johansson} P.~H.,  {Naab} T.,   {Burkert} A.,  2009, \mn@doi [\apj]
  {10.1088/0004-637X/690/1/802}, \href
  {https://ui.adsabs.harvard.edu/abs/2009ApJ...690..802J} {690, 802}

\bibitem[\protect\citeauthoryear{{Knapen}, {Cisternas}  \&
  {Querejeta}}{{Knapen} et~al.}{2015}]{Knapen2015}
{Knapen} J.~H.,  {Cisternas} M.,   {Querejeta} M.,  2015, \mn@doi [\mnras]
  {10.1093/mnras/stv2135}, \href
  {https://ui.adsabs.harvard.edu/abs/2015MNRAS.454.1742K} {454, 1742}

\bibitem[\protect\citeauthoryear{{Kocevski} et~al.,}{{Kocevski}
  et~al.}{2012}]{Kocevski2012}
{Kocevski} D.~D.,  et~al., 2012, \mn@doi [\apj] {10.1088/0004-637X/744/2/148},
  \href {https://ui.adsabs.harvard.edu/abs/2012ApJ...744..148K} {744, 148}

\bibitem[\protect\citeauthoryear{{Koudmani}, {Somerville}, {Sijacki}, {Bourne},
  {Jiang}  \& {Profit}}{{Koudmani} et~al.}{2024}]{Koudmani2024}
{Koudmani} S.,  {Somerville} R.~S.,  {Sijacki} D.,  {Bourne} M.~A.,  {Jiang}
  Y.-F.,   {Profit} K.,  2024, \mn@doi [\mnras] {10.1093/mnras/stae1422}, \href
  {https://ui.adsabs.harvard.edu/abs/2024MNRAS.532...60K} {532, 60}

\bibitem[\protect\citeauthoryear{{La Marca} et~al.,}{{La Marca}
  et~al.}{2024}]{LaMarca2024}
{La Marca} A.,  et~al., 2024, \mn@doi [\aap] {10.1051/0004-6361/202348188},
  \href {https://ui.adsabs.harvard.edu/abs/2024A&A...690A.326L} {690, A326}

\bibitem[\protect\citeauthoryear{{Lambrides} et~al.,}{{Lambrides}
  et~al.}{2021}]{Lambrides2021_classification}
{Lambrides} E.~L.,  et~al., 2021, \mn@doi [\apj] {10.3847/1538-4357/ac0fdf},
  \href {https://ui.adsabs.harvard.edu/abs/2021ApJ...919...43L} {919, 43}

\bibitem[\protect\citeauthoryear{{Li} et~al.,}{{Li} et~al.}{2023a}]{Li2023psb}
{Li} W.,  et~al., 2023a, \mn@doi [\mnras] {10.1093/mnras/stad1473}, \href
  {https://ui.adsabs.harvard.edu/abs/2023MNRAS.523..720L} {523, 720}

\bibitem[\protect\citeauthoryear{{Li} et~al.,}{{Li} et~al.}{2023b}]{Li2023}
{Li} W.,  et~al., 2023b, \mn@doi [\apj] {10.3847/1538-4357/acb13d}, \href
  {https://ui.adsabs.harvard.edu/abs/2023ApJ...944..168L} {944, 168}

\bibitem[\protect\citeauthoryear{{Liao}, {Irodotou}, {Johansson}, {Naab},
  {Rizzuto}, {Hislop}, {Rawlings}  \& {Wright}}{{Liao} et~al.}{2024}]{Liao2024}
{Liao} S.,  {Irodotou} D.,  {Johansson} P.~H.,  {Naab} T.,  {Rizzuto} F.~P.,
  {Hislop} J.~M.,  {Rawlings} A.,   {Wright} R.~J.,  2024, \mn@doi [\mnras]
  {10.1093/mnras/stae360}, \href
  {https://ui.adsabs.harvard.edu/abs/2024MNRAS.528.5080L} {528, 5080}

\bibitem[\protect\citeauthoryear{{Lotz}, {Jonsson}, {Cox}  \& {Primack}}{{Lotz}
  et~al.}{2008}]{Lotz2008}
{Lotz} J.~M.,  {Jonsson} P.,  {Cox} T.~J.,   {Primack} J.~R.,  2008, \mn@doi
  [\mnras] {10.1111/j.1365-2966.2008.14004.x}, \href
  {https://ui.adsabs.harvard.edu/abs/2008MNRAS.391.1137L} {391, 1137}

\bibitem[\protect\citeauthoryear{{Lotz}, {Jonsson}, {Cox}  \& {Primack}}{{Lotz}
  et~al.}{2010}]{Lotz2010b}
{Lotz} J.~M.,  {Jonsson} P.,  {Cox} T.~J.,   {Primack} J.~R.,  2010, \mn@doi
  [\mnras] {10.1111/j.1365-2966.2010.16268.x}, \href
  {https://ui.adsabs.harvard.edu/abs/2010MNRAS.404..575L} {404, 575}

\bibitem[\protect\citeauthoryear{{Marian} et~al.,}{{Marian}
  et~al.}{2019}]{Marian2019}
{Marian} V.,  et~al., 2019, \mn@doi [\apj] {10.3847/1538-4357/ab385b}, \href
  {https://ui.adsabs.harvard.edu/abs/2019ApJ...882..141M} {882, 141}

\bibitem[\protect\citeauthoryear{Marinacci et~al.,}{Marinacci
  et~al.}{2018}]{Mariancci2018}
Marinacci F.,  et~al., 2018, \mn@doi [\mnras] {10.1093/mnras/sty2206}, 480,
  5113

\bibitem[\protect\citeauthoryear{{McAlpine}, {Harrison}, {Rosario},
  {Alexander}, {Ellison}, {Johansson}  \& {Patton}}{{McAlpine}
  et~al.}{2020}]{McAlpine2020}
{McAlpine} S.,  {Harrison} C.~M.,  {Rosario} D.~J.,  {Alexander} D.~M.,
  {Ellison} S.~L.,  {Johansson} P.~H.,   {Patton} D.~R.,  2020, \mn@doi
  [\mnras] {10.1093/mnras/staa1123}, \href
  {https://ui.adsabs.harvard.edu/abs/2020MNRAS.494.5713M} {494, 5713}

\bibitem[\protect\citeauthoryear{{McElroy} et~al.,}{{McElroy}
  et~al.}{2022}]{McElroy2022}
{McElroy} R.,  et~al., 2022, \mn@doi [\mnras] {10.1093/mnras/stac1715}, \href
  {https://ui.adsabs.harvard.edu/abs/2022MNRAS.515.3406M} {515, 3406}

\bibitem[\protect\citeauthoryear{{Mechtley} et~al.,}{{Mechtley}
  et~al.}{2016}]{Mechtley2016}
{Mechtley} M.,  et~al., 2016, \mn@doi [\apj] {10.3847/0004-637X/830/2/156},
  \href {https://ui.adsabs.harvard.edu/abs/2016ApJ...830..156M} {830, 156}

\bibitem[\protect\citeauthoryear{Naiman et~al.,}{Naiman
  et~al.}{2018}]{Naiman2018}
Naiman J.~P.,  et~al., 2018, \mn@doi [\mnras] {10.1093/mnras/sty618}, 477, 1206

\bibitem[\protect\citeauthoryear{Nelson et~al.,}{Nelson
  et~al.}{2017}]{Nelson2017}
Nelson D.,  et~al., 2017, \mn@doi [\mnras] {10.1093/mnras/stx3040}, 475, 624

\bibitem[\protect\citeauthoryear{{Nelson} et~al.,}{{Nelson}
  et~al.}{2019a}]{Nelson2019b}
{Nelson} D.,  et~al., 2019a, \mn@doi [Computational Astrophysics and Cosmology]
  {10.1186/s40668-019-0028-x}, \href
  {https://ui.adsabs.harvard.edu/abs/2019ComAC...6....2N} {6, 2}

\bibitem[\protect\citeauthoryear{{Nelson} et~al.,}{{Nelson}
  et~al.}{2019b}]{Nelson2019}
{Nelson} D.,  et~al., 2019b, \mn@doi [\mnras] {10.1093/mnras/stz2306}, \href
  {https://ui.adsabs.harvard.edu/abs/2019MNRAS.490.3234N} {490, 3234}

\bibitem[\protect\citeauthoryear{{Omori} et~al.,}{{Omori}
  et~al.}{2023}]{Omori2023}
{Omori} K.~C.,  et~al., 2023, \mn@doi [\aap] {10.1051/0004-6361/202346743},
  \href {https://ui.adsabs.harvard.edu/abs/2023A&A...679A.142O} {679, A142}

\bibitem[\protect\citeauthoryear{{Patton}, {Torrey}, {Ellison}, {Mendel}  \&
  {Scudder}}{{Patton} et~al.}{2013}]{Patton2013}
{Patton} D.~R.,  {Torrey} P.,  {Ellison} S.~L.,  {Mendel} J.~T.,   {Scudder}
  J.~M.,  2013, \mn@doi [\mnras] {10.1093/mnrasl/slt058}, \href
  {https://ui.adsabs.harvard.edu/abs/2013MNRAS.433L..59P} {433, L59}

\bibitem[\protect\citeauthoryear{Patton et~al.,}{Patton
  et~al.}{2020}]{Patton2020}
Patton D.~R.,  et~al., 2020, \mn@doi [\mnras] {10.1093/mnras/staa913}, 494,
  4969

\bibitem[\protect\citeauthoryear{{Patton}, {Faria}, {Hani}, {Torrey},
  {Ellison}, {Thakur}  \& {Westlake}}{{Patton} et~al.}{2024}]{Patton2024}
{Patton} D.~R.,  {Faria} L.,  {Hani} M.~H.,  {Torrey} P.,  {Ellison} S.~L.,
  {Thakur} S.~D.,   {Westlake} R.~I.,  2024, \mn@doi [\mnras]
  {10.1093/mnras/stae608}, \href
  {https://ui.adsabs.harvard.edu/abs/2024MNRAS.529.1493P} {529, 1493}

\bibitem[\protect\citeauthoryear{{Pearson}, {Wang}, {Trayford}, {Petrillo}  \&
  {van der Tak}}{{Pearson} et~al.}{2019}]{Pearson2019}
{Pearson} W.~J.,  {Wang} L.,  {Trayford} J.~W.,  {Petrillo} C.~E.,   {van der
  Tak} F.~F.~S.,  2019, \mn@doi [\aap] {10.1051/0004-6361/201935355}, \href
  {https://ui.adsabs.harvard.edu/abs/2019A&A...626A..49P} {626, A49}

\bibitem[\protect\citeauthoryear{{Pierce} et~al.,}{{Pierce}
  et~al.}{2022}]{Pierce2022}
{Pierce} J.~C.~S.,  et~al., 2022, \mn@doi [\mnras] {10.1093/mnras/stab3231},
  \href {https://ui.adsabs.harvard.edu/abs/2022MNRAS.510.1163P} {510, 1163}

\bibitem[\protect\citeauthoryear{Pillepich et~al.,}{Pillepich
  et~al.}{2017}]{Pillepich20171}
Pillepich A.,  et~al., 2017, \mn@doi [\mnras] {10.1093/mnras/stx3112}, 475, 648

\bibitem[\protect\citeauthoryear{{Pillepich} et~al.,}{{Pillepich}
  et~al.}{2018}]{Pillepich2018}
{Pillepich} A.,  et~al., 2018, \mn@doi [\mnras] {10.1093/mnras/stx2656}, \href
  {https://ui.adsabs.harvard.edu/abs/2018MNRAS.473.4077P} {473, 4077}

\bibitem[\protect\citeauthoryear{{Pillepich} et~al.,}{{Pillepich}
  et~al.}{2019}]{Pillepich2019}
{Pillepich} A.,  et~al., 2019, \mn@doi [\mnras] {10.1093/mnras/stz2338}, \href
  {https://ui.adsabs.harvard.edu/abs/2019MNRAS.490.3196P} {490, 3196}

\bibitem[\protect\citeauthoryear{Quai, Hani, Ellison, Patton  \& Woo}{Quai
  et~al.}{2021}]{Quai2021}
Quai S.,  Hani M.~H.,  Ellison S.~L.,  Patton D.~R.,   Woo J.,  2021, \mn@doi
  [\mnras] {10.1093/mnras/stab988}, 504, 1888

\bibitem[\protect\citeauthoryear{{Quai}, {Byrne-Mamahit}, {Ellison}, {Patton}
  \& {Hani}}{{Quai} et~al.}{2023}]{Quai2023}
{Quai} S.,  {Byrne-Mamahit} S.,  {Ellison} S.~L.,  {Patton} D.~R.,   {Hani}
  M.~H.,  2023, \mn@doi [\mnras] {10.1093/mnras/stac3713}, \href
  {https://ui.adsabs.harvard.edu/abs/2023MNRAS.519.2119Q} {519, 2119}

\bibitem[\protect\citeauthoryear{Rodriguez-Gomez et~al.,}{Rodriguez-Gomez
  et~al.}{2015}]{Rodriguez-Gomez2015}
Rodriguez-Gomez V.,  et~al., 2015, \mn@doi [\mnras] {10.1093/mnras/stv264},
  449, 49

\bibitem[\protect\citeauthoryear{{Satyapal}, {Ellison}, {McAlpine}, {Hickox},
  {Patton}  \& {Mendel}}{{Satyapal} et~al.}{2014}]{Satyapal2014}
{Satyapal} S.,  {Ellison} S.~L.,  {McAlpine} W.,  {Hickox} R.~C.,  {Patton}
  D.~R.,   {Mendel} J.~T.,  2014, \mn@doi [\mnras] {10.1093/mnras/stu650},
  \href {https://ui.adsabs.harvard.edu/abs/2014MNRAS.441.1297S} {441, 1297}

\bibitem[\protect\citeauthoryear{{Schawinski}, {Treister}, {Urry}, {Cardamone},
  {Simmons}  \& {Yi}}{{Schawinski} et~al.}{2011}]{Schawinski2011}
{Schawinski} K.,  {Treister} E.,  {Urry} C.~M.,  {Cardamone} C.~N.,  {Simmons}
  B.,   {Yi} S.~K.,  2011, \mn@doi [\apjl] {10.1088/2041-8205/727/2/L31}, \href
  {https://ui.adsabs.harvard.edu/abs/2011ApJ...727L..31S} {727, L31}

\bibitem[\protect\citeauthoryear{{Schawinski}, {Simmons}, {Urry}, {Treister}
  \& {Glikman}}{{Schawinski} et~al.}{2012}]{Schawinski2012}
{Schawinski} K.,  {Simmons} B.~D.,  {Urry} C.~M.,  {Treister} E.,   {Glikman}
  E.,  2012, \mn@doi [\mnras] {10.1111/j.1745-3933.2012.01302.x}, \href
  {https://ui.adsabs.harvard.edu/abs/2012MNRAS.425L..61S} {425, L61}

\bibitem[\protect\citeauthoryear{{Schechter} et~al.,}{{Schechter}
  et~al.}{2025}]{Schechter2025}
{Schechter} A.~L.,  et~al., 2025, \mn@doi [arXiv e-prints]
  {10.48550/arXiv.2507.01092}, \href
  {https://ui.adsabs.harvard.edu/abs/2025arXiv250701092S} {p. arXiv:2507.01092}

\bibitem[\protect\citeauthoryear{{Scudder}, {Ellison}, {Torrey}, {Patton}  \&
  {Mendel}}{{Scudder} et~al.}{2012}]{Scudder2012}
{Scudder} J.~M.,  {Ellison} S.~L.,  {Torrey} P.,  {Patton} D.~R.,   {Mendel}
  J.~T.,  2012, \mn@doi [\mnras] {10.1111/j.1365-2966.2012.21749.x}, \href
  {https://ui.adsabs.harvard.edu/abs/2012MNRAS.426..549S} {426, 549}

\bibitem[\protect\citeauthoryear{{Secrest}, {Ellison}, {Satyapal}  \&
  {Blecha}}{{Secrest} et~al.}{2020}]{Secrest2020}
{Secrest} N.~J.,  {Ellison} S.~L.,  {Satyapal} S.,   {Blecha} L.,  2020,
  \mn@doi [\mnras] {10.1093/mnras/staa1692}, \href
  {https://ui.adsabs.harvard.edu/abs/2020MNRAS.499.2380S} {499, 2380}

\bibitem[\protect\citeauthoryear{{Shah} et~al.,}{{Shah}
  et~al.}{2020}]{Shah2020}
{Shah} E.~A.,  et~al., 2020, \mn@doi [\apj] {10.3847/1538-4357/abbf59}, \href
  {https://ui.adsabs.harvard.edu/abs/2020ApJ...904..107S} {904, 107}

\bibitem[\protect\citeauthoryear{{Silva}, {Marchesini}, {Silverman}, {Martis},
  {Iono}, {Espada}  \& {Skelton}}{{Silva} et~al.}{2021}]{Silva2021}
{Silva} A.,  {Marchesini} D.,  {Silverman} J.~D.,  {Martis} N.,  {Iono} D.,
  {Espada} D.,   {Skelton} R.,  2021, \mn@doi [\apj]
  {10.3847/1538-4357/abdbb1}, \href
  {https://ui.adsabs.harvard.edu/abs/2021ApJ...909..124S} {909, 124}

\bibitem[\protect\citeauthoryear{{Silverman} et~al.,}{{Silverman}
  et~al.}{2011}]{Silverman2011}
{Silverman} J.~D.,  et~al., 2011, \mn@doi [\apj] {10.1088/0004-637X/743/1/2},
  \href {https://ui.adsabs.harvard.edu/abs/2011ApJ...743....2S} {743, 2}

\bibitem[\protect\citeauthoryear{{Sotillo-Ramos} et~al.,}{{Sotillo-Ramos}
  et~al.}{2022}]{SotilloRamos2022}
{Sotillo-Ramos} D.,  et~al., 2022, \mn@doi [\mnras] {10.1093/mnras/stac2586},
  \href {https://ui.adsabs.harvard.edu/abs/2022MNRAS.516.5404S} {516, 5404}

\bibitem[\protect\citeauthoryear{{Springel}}{{Springel}}{2010}]{Springel2010}
{Springel} V.,  2010, \mn@doi [\mnras] {10.1111/j.1365-2966.2009.15715.x},
  \href {https://ui.adsabs.harvard.edu/abs/2010MNRAS.401..791S} {401, 791}

\bibitem[\protect\citeauthoryear{{Springel}, {Di Matteo}  \&
  {Hernquist}}{{Springel} et~al.}{2005}]{Springel2005}
{Springel} V.,  {Di Matteo} T.,   {Hernquist} L.,  2005, \mn@doi [\mnras]
  {10.1111/j.1365-2966.2005.09238.x}, \href
  {https://ui.adsabs.harvard.edu/abs/2005MNRAS.361..776S} {361, 776}

\bibitem[\protect\citeauthoryear{Springel et~al.,}{Springel
  et~al.}{2017}]{Springel2017}
Springel V.,  et~al., 2017, \mn@doi [\mnras] {10.1093/mnras/stx3304}, 475, 676

\bibitem[\protect\citeauthoryear{{Steinborn}, {Hirschmann}, {Dolag}, {Shankar},
  {Juneau}, {Krumpe}, {Remus}  \& {Teklu}}{{Steinborn}
  et~al.}{2018}]{Steinborn2018}
{Steinborn} L.~K.,  {Hirschmann} M.,  {Dolag} K.,  {Shankar} F.,  {Juneau} S.,
  {Krumpe} M.,  {Remus} R.-S.,   {Teklu} A.~F.,  2018, \mn@doi [\mnras]
  {10.1093/mnras/sty2288}, \href
  {https://ui.adsabs.harvard.edu/abs/2018MNRAS.481..341S} {481, 341}

\bibitem[\protect\citeauthoryear{Terrazas et~al.,}{Terrazas
  et~al.}{2020}]{Terrazas2020}
Terrazas B.~A.,  et~al., 2020, \mn@doi [\mnras] {10.1093/mnras/staa374}, 493,
  1888

\bibitem[\protect\citeauthoryear{{Thorp}, {Ellison}, {Simard}, {S{\'a}nchez}
  \& {Antonio}}{{Thorp} et~al.}{2019}]{Thorp2019}
{Thorp} M.~D.,  {Ellison} S.~L.,  {Simard} L.,  {S{\'a}nchez} S.~F.,
  {Antonio} B.,  2019, \mn@doi [\mnras] {10.1093/mnrasl/sly185}, \href
  {https://ui.adsabs.harvard.edu/abs/2019MNRAS.482L..55T} {482, L55}

\bibitem[\protect\citeauthoryear{{Thorp}, {Bluck}, {Ellison}, {Maiolino},
  {Conselice}, {Hani}  \& {Bottrell}}{{Thorp} et~al.}{2021}]{Thorp2021}
{Thorp} M.~D.,  {Bluck} A. F.~L.,  {Ellison} S.~L.,  {Maiolino} R.,
  {Conselice} C.~J.,  {Hani} M.~H.,   {Bottrell} C.,  2021, \mn@doi [\mnras]
  {10.1093/mnras/stab2201}, \href
  {https://ui.adsabs.harvard.edu/abs/2021MNRAS.507..886T} {507, 886}

\bibitem[\protect\citeauthoryear{{Treister}, {Schawinski}, {Urry}  \&
  {Simmons}}{{Treister} et~al.}{2012}]{Treister2012}
{Treister} E.,  {Schawinski} K.,  {Urry} C.~M.,   {Simmons} B.~D.,  2012,
  \mn@doi [\apjl] {10.1088/2041-8205/758/2/L39}, \href
  {https://ui.adsabs.harvard.edu/abs/2012ApJ...758L..39T} {758, L39}

\bibitem[\protect\citeauthoryear{{Urrutia}, {Lacy}  \& {Becker}}{{Urrutia}
  et~al.}{2008}]{Urrutia2008}
{Urrutia} T.,  {Lacy} M.,   {Becker} R.~H.,  2008, \mn@doi [\apj]
  {10.1086/523959}, \href
  {https://ui.adsabs.harvard.edu/abs/2008ApJ...674...80U} {674, 80}

\bibitem[\protect\citeauthoryear{{V{\'a}squez-Bustos}, {Argudo-Fern{\'a}ndez},
  {Boquien}, {Castillo-Baeza}, {Castillo-Rencoret}  \&
  {Ariza-Quintana}}{{V{\'a}squez-Bustos} et~al.}{2025}]{VasquezBustos2025}
{V{\'a}squez-Bustos} P.,  {Argudo-Fern{\'a}ndez} M.,  {Boquien} M.,
  {Castillo-Baeza} N.,  {Castillo-Rencoret} A.,   {Ariza-Quintana} D.,  2025,
  \mn@doi [\aap] {10.1051/0004-6361/202451464}, \href
  {https://ui.adsabs.harvard.edu/abs/2025A&A...696A.206V} {696, A206}

\bibitem[\protect\citeauthoryear{{Villforth}}{{Villforth}}{2023}]{Villforth2023}
{Villforth} C.,  2023, \mn@doi [arXiv e-prints] {10.48550/arXiv.2309.03276},
  \href {https://ui.adsabs.harvard.edu/abs/2023arXiv230903276V} {p.
  arXiv:2309.03276}

\bibitem[\protect\citeauthoryear{{Villforth} et~al.,}{{Villforth}
  et~al.}{2014}]{Villforth2014}
{Villforth} C.,  et~al., 2014, \mn@doi [\mnras] {10.1093/mnras/stu173}, \href
  {https://ui.adsabs.harvard.edu/abs/2014MNRAS.439.3342V} {439, 3342}

\bibitem[\protect\citeauthoryear{{Villforth} et~al.,}{{Villforth}
  et~al.}{2017}]{Villforth2017}
{Villforth} C.,  et~al., 2017, \mn@doi [\mnras] {10.1093/mnras/stw3037}, \href
  {https://ui.adsabs.harvard.edu/abs/2017MNRAS.466..812V} {466, 812}

\bibitem[\protect\citeauthoryear{{Weinberger} et~al.,}{{Weinberger}
  et~al.}{2017}]{Weinberger2017}
{Weinberger} R.,  et~al., 2017, \mn@doi [\mnras] {10.1093/mnras/stw2944}, \href
  {https://ui.adsabs.harvard.edu/abs/2017MNRAS.465.3291W} {465, 3291}

\bibitem[\protect\citeauthoryear{{Weinberger}, {Bhowmick}, {Blecha}, {Bryan},
  {Buchner}, {Hernquist}, {Hlavacek-Larrondo}  \& {Springel}}{{Weinberger}
  et~al.}{2025}]{Weinberger2025}
{Weinberger} R.,  {Bhowmick} A.,  {Blecha} L.,  {Bryan} G.,  {Buchner} J.,
  {Hernquist} L.,  {Hlavacek-Larrondo} J.,   {Springel} V.,  2025, \mn@doi
  [arXiv e-prints] {10.48550/arXiv.2502.13241}, \href
  {https://ui.adsabs.harvard.edu/abs/2025arXiv250213241W} {p. arXiv:2502.13241}

\bibitem[\protect\citeauthoryear{{Weston}, {McIntosh}, {Brodwin}, {Mann},
  {Cooper}, {McConnell}  \& {Nielsen}}{{Weston} et~al.}{2017}]{Weston2017}
{Weston} M.~E.,  {McIntosh} D.~H.,  {Brodwin} M.,  {Mann} J.,  {Cooper} A.,
  {McConnell} A.,   {Nielsen} J.~L.,  2017, \mn@doi [\mnras]
  {10.1093/mnras/stw2620}, \href
  {https://ui.adsabs.harvard.edu/abs/2017MNRAS.464.3882W} {464, 3882}

\bibitem[\protect\citeauthoryear{{White} \& {Rees}}{{White} \&
  {Rees}}{1978}]{WhiteReese1978}
{White} S.~D.~M.,  {Rees} M.~J.,  1978, \mn@doi [\mnras]
  {10.1093/mnras/183.3.341}, \href
  {https://ui.adsabs.harvard.edu/abs/1978MNRAS.183..341W} {183, 341}

\bibitem[\protect\citeauthoryear{{Wilkinson}, {Ellison}, {Bottrell}, {Bickley},
  {Byrne-Mamahit}, {Ferreira}  \& {Patton}}{{Wilkinson}
  et~al.}{2024}]{Wilkinson2024}
{Wilkinson} S.,  {Ellison} S.~L.,  {Bottrell} C.,  {Bickley} R.~W.,
  {Byrne-Mamahit} S.,  {Ferreira} L.,   {Patton} D.~R.,  2024, \mn@doi [\mnras]
  {10.1093/mnras/stae287}

\bibitem[\protect\citeauthoryear{{Woods} \& {Geller}}{{Woods} \&
  {Geller}}{2007}]{Woods2007}
{Woods} D.~F.,  {Geller} M.~J.,  2007, \mn@doi [\aj] {10.1086/519381}, \href
  {https://ui.adsabs.harvard.edu/abs/2007AJ....134..527W} {134, 527}

\bibitem[\protect\citeauthoryear{{Woods}, {Geller}  \& {Barton}}{{Woods}
  et~al.}{2006}]{Woods2006}
{Woods} D.~F.,  {Geller} M.~J.,   {Barton} E.~J.,  2006, \mn@doi [\aj]
  {10.1086/504834}, \href
  {https://ui.adsabs.harvard.edu/abs/2006AJ....132..197W} {132, 197}

\bibitem[\protect\citeauthoryear{{Woods}, {Geller}, {Kurtz}, {Westra},
  {Fabricant}  \& {Dell'Antonio}}{{Woods} et~al.}{2010}]{Woods2010}
{Woods} D.~F.,  {Geller} M.~J.,  {Kurtz} M.~J.,  {Westra} E.,  {Fabricant}
  D.~G.,   {Dell'Antonio} I.,  2010, \mn@doi [\aj]
  {10.1088/0004-6256/139/5/1857}, \href
  {https://ui.adsabs.harvard.edu/abs/2010AJ....139.1857W} {139, 1857}

\bibitem[\protect\citeauthoryear{{Wu}, {Zhu}, {Chang}, {Zeng}  \& {Lei}}{{Wu}
  et~al.}{2025}]{Wu2025}
{Wu} X.,  {Zhu} L.,  {Chang} J.,  {Zeng} G.,   {Lei} Y.,  2025, \mn@doi [\aap]
  {10.1051/0004-6361/202554638}, \href
  {https://ui.adsabs.harvard.edu/abs/2025A&A...699A.374W} {699, A374}

\bibitem[\protect\citeauthoryear{{Yang}, {Ge}  \& {Lu}}{{Yang}
  et~al.}{2019}]{Yang2019}
{Yang} C.,  {Ge} J.-Q.,   {Lu} Y.-J.,  2019, \mn@doi [Research in Astronomy and
  Astrophysics] {10.1088/1674-4527/19/12/177}, \href
  {https://ui.adsabs.harvard.edu/abs/2019RAA....19..177Y} {19, 177}

\bibitem[\protect\citeauthoryear{{Zeng}, {Wang}  \& {Gao}}{{Zeng}
  et~al.}{2021}]{Zeng2021}
{Zeng} G.,  {Wang} L.,   {Gao} L.,  2021, \mn@doi [\mnras]
  {10.1093/mnras/stab2294}, \href
  {https://ui.adsabs.harvard.edu/abs/2021MNRAS.507.3301Z} {507, 3301}

\makeatother
\end{thebibliography}






\bsp	
\label{lastpage}
\end{document}